\documentclass[traditabstract]{aa}

\usepackage{graphicx}

\newcommand{\kms}{km\,s$^{-1}$}
\newcommand{\Mdot}{\mbox{$\dot M$}}                 
\newcommand{\Msolyr}{~M$_{\odot}$\,yr$^{-1}$}      
\newcommand{\Lsol}{\mbox{L$_{\odot}$}}              
\newcommand{\Msol}{\mbox{M$_{\odot}$}}              
\newcommand{\Rsol}{\mbox{R$_{\odot}$}}              

\newcommand{\NaI}{\ion{Na}{i}}
\newcommand{\KI}{\ion{K}{i}}
\newcommand{\HI}{\ion{H}{i}}
\newcommand{\HII}{\ion{H}{ii}}
\newcommand{\CI}{\ion{C}{i}}

\begin{document}

\title{The mass-loss rates of red supergiants and the de Jager prescription}

\author{N. Mauron\inst{1} \and E. Josselin\inst{1}}

\offprints{N.~Mauron}

\institute{GRAAL, CNRS and Universit\'e  Montpellier II,  
 Place Bataillon, 34095 Montpellier, France\\ 
 \email{nicolas.mauron@univ-montp2.fr} }
 

\date{}
\abstract{Mass loss of red supergiants (RSG) is important for the evolution of massive stars,
but is not fully explained. Several empirical prescriptions have been proposed, trying to express 
the mass-loss rate ($\Mdot$) as a function of fundamental stellar parameters (mass, luminosity,
effective temperature). Our goal is to test whether the de Jager  et al.\,\,(1988) prescription,
used in some stellar evolution models, is still valid in view of more recent mass-loss
determinations. By considering 40 Galactic RSGs presenting an infrared excess and an IRAS
 60-$\mu$m flux larger than 2 Jy, and assuming a gas-to-dust mass ratio of 200,
 it is found that that the de Jager rate agrees within a factor 4 with
most $\Mdot$ estimates based on the 60-$\mu$m signal. It is also in agreement 
with 6 of the only 8 Galactic RSGs for which $\Mdot$\ can be measured more directly 
through observations of the circumstellar gas.  The two objects that do not follow the de Jager
prescription (by an order of magnitude) are $\mu$\ Cep and NML Cyg.  We have
also considered the RSGs of the Magellanic Clouds. Thanks to the works of Groenewegen 
et al.\,\,(2009) and Bonanos et al.\,\,(2010), we find that the  RSGs of the Small Magellanic Cloud 
 have mass-loss rates consistent
with the de Jager rate  scaled by $(Z/Z_{\sun})^{\alpha}$, where $Z$ is the metallicity and
$\alpha$ is  0.7. The situation is less clear for the  RSGs of the Large Magellanic Cloud. 
In particular, for $L > 1.6 \times 10^{5}$\ \Lsol, one finds numerous RSGs (except WOH-G64) having
$\Mdot$\  significantly smaller  than the de Jager rate and indicating that $\Mdot$  would no longer
increase with $L$. Before this odd situation is confirmed through further analysis of LMC RSGs, 
we suggest to keep the de Jager prescription unchanged at solar metallicity
in the stellar evolutionary models and to apply a $(Z/Z_{\sun})^{0.7}$  dependence.}

\keywords{Stars: massive;  Stars: mass-loss; Supergiants } 
\titlerunning{The mass-loss rates of red supergiants}
\authorrunning{}
\maketitle


\section{Introduction}

The red supergiant stars (hereafter RSG) are massive stars evolving beyond 
the main sequence and passing a fraction of their lives in the cool 
upper region of the Hertzsprung-Russell (H-R) luminosity-temperature diagram. 
In the solar neighborhood, the initial mass of RSGs is about 8 to 25 \Msol,
their effective temperature is for most objects 3000 to 4000K, and their
luminosity is between 2~$\times$~10$^4$ and $\sim 6$~$\times$~10$^{5}$ \Lsol.
Several  important  works have recently revisited the basic properties of these 
stars in the Galaxy and in other systems like  the  Magellanic Clouds and M31
(Massey 1998; Levesque et al.\,\,2005, hereafter LM05;  Massey et al.\,\,2008; 
 Levesque 2009 and references therein).
In general, a rather good agreement between recent observations and the
predictions of evolutionary models has been established. However, one of the
important processes occuring at the surface of these stars, mass loss, is still
poorly known. The radii of RSGs are huge, typically 500-1500\,\Rsol. 
Consequently, the gravity
of these stars is very small, about 50\,000 times smaller than that of the Sun.
The small gravity favours ejection of matter to the interstellar medium, and
indeed some RSGs lose mass at a high rate, 10$^{-6}$ to 10$^{-4}$ \Msolyr.

Mass loss has several consequences. First, a significant fraction of the 
stellar mass is lost, which is important for subsequent evolution of the star.
Secondly, the ejected circumstellar matter enriches the ISM in dust, e.g. 
diverse types of  silicates  (see for example Verhoelst et al.\,\,2009) and 
even carbonaceous matter (Sylvester et al.\,\,1994, 1998). One can note also 
that some  types of core-collapse supernovae have RSG progenitors (Smartt 
et al.\,\,2009; Josselin \& Lan\c{c}on\,2009), 
so that the density and radial distribution of the RSG winds play 
an important role in the  expansion and evolution of the supernova remnant
(Chevalier et al.\,\,2006;  Smith et al.\,\,2009; van Loon\,\,2010).

 Despite the importance of RSG mass loss, a quantitative physical model
of the winds has yet to be established. Like  for other mass-losing cool giants, 
atmospheric motions such as  pulsation, convection and shocks occur 
(e.g. Kiss et al.\,\,2006; Josselin \& Plez 2007),  
and this may produce  out-moving gas sufficiently cool for dust to form
Then, radiation pressure  on the grains may be sufficient to drive the 
outflow (see for example Habing et al.\,\,1994). Although this scenario is plausible, 
some important details are still unclear. For example, the condensation of the  
Fe-rich silicate grains, crucial to drive the wind, is not well understood
(Woitke 2007). When dust is incompletely formed or absent, other mass-loss 
mechanisms have to be found, but none is completely satisfactory 
(e.g. Bennett 2010 and references therein).

Because no physical model exists,
diverse empirical parametrizations are used in stellar evolution codes 
for estimating \Mdot\, as a function of  basic stellar  characteristics, 
i.e. mass, luminosity, and effective temperature. 
 An important example is the empirical law built by  de Jager et 
al.\,\,(1988; hereafter dJ88). 
This law was based on observational estimates of  \Mdot\, for stars
located over the whole H-R diagram, including 15 RSGs. It is used in the
Cambridge stellar evolution code with a (Z/Z$_{\sun}$)$^{0.5}$ metallicity scaling 
(Stancliffe \& Eldridge 2009). The dJ88 law
 was also used in past evolution models of the Geneva group 
(described in, e.g., Maeder \& Meynet 1989; Eggenberger et al.\,\,2008). However, 
in  the most recent Geneva code, the dJ88 prescription is used only for 
yellow supergiants with $T_{\rm eff}$ between 5000\,K and 8000\,K.
For smaller $T_{\rm eff}$, a power-law in $L$, $\Mdot \propto L^{1.7}$,  is adopted
with no metallicity scaling.

The goal of this paper is 
to investigate whether this de Jager formulation is still in agreement 
with mass-loss rates obtained more  recently. For example, 
mass-loss rates of RSGs were estimated  with IRAS observations of the 
circumstellar dust (Jura \& Kleinman 1990, hereafter JK90;  
Josselin et al.\,\,2000, hereafter JB00)
 or from ISO spectra 
(Verhoelst et al.\,\,2009). Several radio millimeter observations of CO 
from RSGs have also been made since 1988 (JB00; 
Woodhams 1993). The question is whether these more recent measurements support 
 the dJ88 prescription. 

 In the following, Sect.~2 recalls the prescription of de Jager and 
compares it with a few others.  In Sect.~3, we compare the dJ88 rate with  
measured mass-loss rates of RSGs, and we consider  three  different cases: 
i)  the RSGs of the  solar neighbourhood and \Mdot\  estimated with their 
infrared excess at 60$\mu$m; ii) the eight cases of these RSGs for which
circumstellar gas has been directly observed and used to derive \Mdot\ ; 
iii) the RSGs of the Magellanic Clouds.
In Sect.~4, we discuss our results and conclude in Sect.~5.

 
 \section{Comparison of de Jager rates with other prescriptions}
 
 In the paper of dJ88, the authors  collected a large number of mass-loss rates 
determined for 271  Galactic stars located over the whole  H-R diagram.
 The mass-loss rates could be matched by a single empirical  formula
 in which the parameters are the effective temperature  $T_{\rm eff}$ and the
 luminosity $L$. Hereafter, we call this
formula ``the de Jager rate''. According to dJ88, the accuracy
 of this formula is about 0.45 in log~\Mdot, i.e. about a factor of 3.
 Because the de Jager   formula comprises 20 Chebycheff polynomia, the dependence
 of \Mdot\, on $T_{\rm eff}$ and $L$ cannot  be seen clearly. Therefore, we show in
 Fig.~1 what is the situation for RSGs. The de Jager rate is plotted
 for two different effective temperatures, 3500 and 4000~K, since the majority
 of Galactic RSGs have their $T_{\rm eff}$  between these values (LM05). 
 More precisely, in the list of 74 RSGs of LM05 (with a M-type classification), 
only $\alpha$\,Her has a lower temperature,  with 3450\,K, but it is not a 
massive star, $M \approx$ 2.1 \Msol\, according to Schr\"{o}der \& Cuntz (2007).
Figure~1 shows that \Mdot\, is not very dependent on $T_{\rm eff}$ in the de Jager 
 formulation, and that \Mdot\, depends only on $L$.

 Among the 271 stars (hot or cool, luminous or not),
which are at the basis for the dJ88 prescription, there are only
15 Galactic RSGs, for which dJ88 compiled $T_{\rm eff}$, $L$ and an observational 
estimate of \Mdot, or an average of estimates. 
These compiled data are listed in Appendix~A, together with results of this work, 
discussed later.  It can be noted that the $T_{\rm eff}$ 
are significantly cooler than those recently determined by LM05, when available. 
In addition,  about half of the estimated \Mdot\ are based on the early paper of 
Gehrz and Woolf\,(1971). So, we think it appropriate
to reexamine the validity of the dJ88 prescription with a larger sample of objects 
(40 are considered in this work)  and more recent data.

  There exists a second formulation  published later by Nieuwenhuijzen and de Jager
 (1990; hereafter NJ90).  The stellar mass is taken into account, and 
the dependence of \Mdot\, on stellar mass, luminosity and
 temperature is more explicit. NJ90 find:

\smallskip
\smallskip    
\noindent
 log $\Mdot$ =\,$-$7.93\,$+$\,1.64\,log $L$\,$+$\,0.16\,log $M$\,$-$\,1.61\,log $T_{\rm eff}$.

\smallskip
\smallskip

It can be seen in the above formula that \Mdot\ depends very little on $M$, since for
$M = $9 \Msol\ and 25 \Msol, the second term is equal to 0.15 and 0.22 respectively.
 This represents a difference of only 17 percent on \Mdot\,, and consequently this
 mass term can be considered to be constant  and equal to 0.18. Then, a more explicit
formulation is:

\smallskip
\smallskip

$\Mdot = 5.6 \times 10^{-6} (L/10^5)^{1.64} \,\,(T_{\rm eff}/3500)^{-1.61} $.

\smallskip
\smallskip

 The  rate of  NJ90
is plotted in Fig.~1 for $T_{\rm eff} = 3750$K.  For other temperatures,
this rate does not change much: for effective temperatures of 3500 or 4000K,
the NJ90 line in Fig~1 would be shifted up or down by only 0.05 in the y-axis. 
This NJ90 is not very different 
from the de Jager rate: the largest difference is about a factor of 2 
at $L \sim 10^{5}$.  NJ90 also note that for cool evolved stars the dJ88 rate
is more accurate than the NJ90 one in fitting the \Mdot\, observational
measurements (their Table 2).

\smallskip  

 We have also plotted in Fig.~1 the rate adopted in the recent Geneva models
(R. Hirschi, private communication). This rate is
represented by a continuous straight line located  very close to,  and just below 
the NJ90 line. This Geneva rate is practically identical to
the  NJ90 one, but there is no dependency on $T_{\rm eff}$. 
The Geneva rate is the following: $\Mdot = 4.7 \times 10^{-6} (L/10^5)^{1.7}$.

\smallskip

\begin{figure}
\includegraphics*[width=7cm,angle=-90]{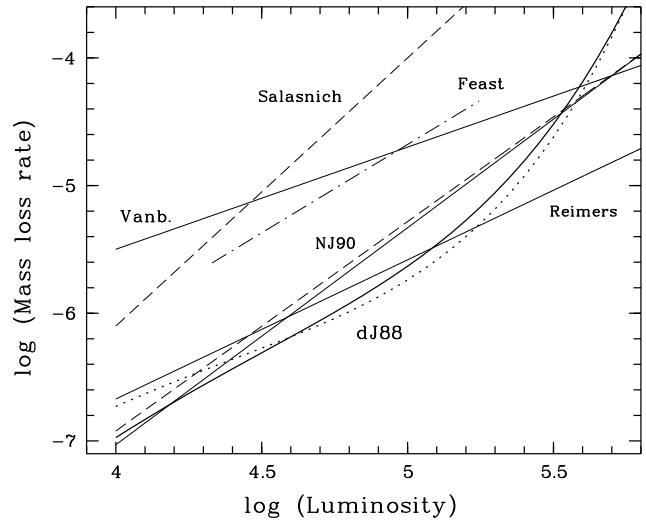}
\caption{ Comparison of several prescriptions for the RSG  mass-loss rates
(unit \Msolyr) plotted against luminosity (in \Lsol). 
The de Jager prescription is represented by  the 2 curved lines 
(solid line $T_{\rm eff} = 4000$\,K, dotted line $T_{\rm eff} = 3500$\,K).}
\end{figure}

Figure~1 also shows the Reimers law (Reimers 1975; Kudritzki \& Reimers 1978):
\smallskip
\smallskip

  $\Mdot = 5.5 \times 10^{-13}$\,$L R / M $, 

\smallskip
\smallskip

\noindent  with $L$, $R$ and $M$ in solar
 units. 
Since we restrict our survey to RSGs, this relation can be simplified: 
$R$ is equal to
 $L^{0.5} \times (T_{\rm eff}/5770)^{-2}$ and the factor depending on $T_{\rm eff}$
does not vary much (2.72 and 2.08 for $T_{\rm eff} = 3500$\, and 4000~K respectively). 
In addition, for the  RSGs studied here,
there is an approximate relation between mass  and luminosity.
For $M = 9$\,\Msol, $L \approx 2.5 \times 10^{4}$\,\Lsol,
and for $M = 25$\,\Msol, $L$\,$\approx 3 \times 10^{5}$\,\Lsol 
(see,  for example, Fig.~8 of Massey et al.~2009, lower panels). 
We find $M$\,$\approx$\,$0.14$\,$L^{0.41}$.  So the Reimers relation 
 for an average temperature $T_{\rm eff} = 3750$K  is:

\smallskip
\smallskip

  $\Mdot = 2.6  \times 10^{-6} (L/10^{5})^{1.09}$\,.

\smallskip
\smallskip

This relation is plotted in Fig.~1, and it can be seen that the Reimers rate 
is close to the dJ88 rate (within a factor of 2 for  $L \leq 2 \times 10^{5}$\,\Lsol), 
but  the latter is significantly stronger for larger luminosities. 
 Again, we have not drawn, for clarity, the Reimers lines corresponding to
other temperatures, but for 3500 and 4000K, they would be 0.060 (in log) above and 
0.056 below the plotted line. So the dependance on temperature is quite small
compared to the observational
 uncertainty on measured mass-loss rates considered in the following 
(typically a factor 2 or 3).

 There exist other mass-loss prescriptions. Feast (1992) noted that there
is a very good correlation between the pulsation period $P$ of 15 RSGs of
the Large Magellanic Cloud and their luminosities $L$ based on JHK and IRAS data and 
estimated by Reid et al.\,\,(1990). The mass-loss rates  published 
by Reid et al.\,\,for these stars also correlate rather well with $P$.
A relation linking $\Mdot$ and $L$ is derivable and given in Equation (12) of
Salasnich et al (1999). This relation (plotted in Fig.~1) can be written:

\smallskip
\smallskip

  $\Mdot = 2.1  \times 10^{-5} (L/10^{5})^{1.385}$\,. 

\smallskip
\smallskip 

 A prescription based also on the LMC RSGs and the work of Reid et al.\,\,(1990) is
used by Vanbeveren et al.\,\,(1998) (see also Vanbeveren et al. 2007). It is labelled
``Vanb.'' in Fig.~1. This rate is in agreement with the Feast rate for 
$L$\,$\sim$ 10$^{5}$\,\Lsol, but its slope is lower. Its expression is:

\smallskip
\smallskip 

 $\Mdot = 2.0  \times 10^{-5} (L/10^{5})^{0.80}$\,.

\smallskip
\smallskip 

  Another mass-loss prescription has been proposed by Salasnich et  al. (1999), who 
adopt the Feast relation, but take a  gas-to-dust  mass ratio increasing with luminosity.
Typically, this ratio would be 200 for AGB stars but could increase to 1000 for
RSGs with a luminosity of 10$^{5}$\,\Lsol. Consequently, the Salasnich prescription
implies very high mass-loss rates.  For $L \sim 10^{5}$\,\Lsol, this prescription
gives  $10^{-4}$\,\Msolyr, about a factor of 50 above the de Jager rate.  The Salasnich 
law is:

\smallskip
\smallskip

 $\Mdot = 1.00  \times 10^{-4} (L/10^{5})^{2.1}$\,. 
\smallskip
\smallskip

 Finally, van Loon et al.\,\,(2005) derived a mass-loss law for AGB stars and RSGs, 
but it cannot be applied if the circumstellar envelope is not very dusty. Nevertheless,
the rate of van Loon et al.\,\,(2005) is interesting to compare with the de Jager rate, and
this will be done  with more details in the discussion. 
In the next section, we  compare the de Jager rate with modern mass-loss measurements.

\section{Comparison of the de Jager rate with observations}

\subsection{The 60-$\mu$m excess emission of RSGs in the solar neighbourhood}

The RSGs of the solar neighbourhood  are key targets for investigating
mass loss, either because they are close to us like Betelgeuse or Antares, or because the
membership to an OB association gives a distance indication and confirms the object to be a 
massive star. These RSGs have been originally listed by Humphreys (1978). JK90
investigated mass loss of the most luminous RSGs. They consider a list of 21 objects
more luminous than 10$^{5}$\,\Lsol\, and estimate mass loss with the IRAS~60$\mu$m flux.
JB00 consider 65  RSGs,  present photometric JHKL measurements 
of most of them, and detect the CO millimeter emission in several cases.
One of their conclusions is a large scatter of \Mdot\ for
a given luminosity, a property that was also noted by JK90.
LM05 have revised  effective temperatures, distances and interstellar
extinction for 74 RSGs. 

 A first list of RSGs was built from the list of 74 objects of LM05, supplemented with
21  objects from JB00. For these 95 sources, we collected
optical (UBVI) and near-infrared photometry (JHKL), preferentially from Lee~(1970). 
Data from  Simbad, 2MASS, JB00 and Gezari et al.\,\,(1993) were also used. 
At mid-infrared wavelengths, IRAS measurements were adopted. 
Then, for most of the stars, photometry from the U band to the 60$\mu$m band is available.
Almost all these stars belong to OB associations, and the distance moduli and interstellar
absorption given by LM05 were adopted. When a star was not examined in the work of LM05,
we took the distance modulus of the association  
from Humphreys~(1978) and the reddening was derived from the averaged reddening of the OB-type
stars of the association. For several objects (e.g., VY CMa, VX Sgr), the distance
and absorption come from recent, individual determinations.

 From this primary list of 95 objects, we are interested 
in those presenting evidence of mass loss.
For that purpose, we selected objects presenting an infrared excess,
measured by the ratio of 12~$\mu$m flux to the dereddened K-band flux.  Following
JK90, only stars with $f_{\rm 12}/f_{K}$ larger than 0.1 were selected. In addition,
we imposed $f_{60} \geq 2$~Jy, in order to work with reliable 60$\mu$m fluxes.
This resulted in a list of 39 objects for which the photometric data are given in 
Appendix~B. The characteristics of these 39 objects are given in Table 1.  An additional 
remarkable RSG, NML Cyg, will be examined in the next subsection.


\begin{table*}[!ht]
\caption[]{Properties of the RSGs. The first 27 stars are from LM05, while the following 12
stars are from JB00. The column marked $V$ is the assumed
wind speed. The column marked $\lambda$ is the flux averaged wavelength. 
 $L_{\rm Lev}$ is the luminosity given in LM05 (from their $M_{\rm bol}$) and 
$L_{\rm phot}$ is the luminosity obtained by integrating the UBVIJHKL + IRAS photometry.
   $\Mdot$ is derived with Jura's formula, where the used luminosity is $L_{\rm phot}$.} 

\begin{center}
\begin{tabular}{lrrrcrrrrrr}
\hline
\hline
\noalign{\smallskip}
Name &  $m-M$ & $E_{\rm B-V}$ & $f_{\rm 12}/ f_{\rm K}$ & $V$\,\,\,  & $D$\,\,\, & $f_{\rm60}$\,\, & $\lambda_m$\,\,\, & $L_{\rm Lev}$\,\,\,\ &  $L_{\rm  phot}$ &$\Mdot$\,\,\,\,\,\,\,\\

\noalign{\smallskip}

 &    (mag)    &  (mag)   &  &     (km\ s$^{-1}$)  & (kpc)  &  (Jy)       & ($\mu$m)  &  (\Lsol)  & (\Lsol)& (\Msolyr)\\

\noalign{\smallskip}
\hline
\noalign{\smallskip}

V589 Cas & 11.50 & 0.78 & 0.156 & 14 & 2.00 &    3.61 & 1.72 &  52\,000 &  35\,000 &  5.0  $\times\ 10^{-7}$\\
BU Per   & 11.40 & 0.66 & 0.435 & 14 & 1.90 &    5.23 & 2.20 &  58\,000 &  38\,000 &  7.4  $\times\ 10^{-7}$\\
SU Per   & 11.40 & 0.66 & 0.241 & 19 & 1.90 &    6.87 & 1.77 &  90\,000 &  85\,000 &  7.7  $\times\ 10^{-7}$\\
RS Per   & 11.90 & 0.56 & 0.417 & 20 & 2.40 &    9.93 & 2.28 & 144\,000 &  95\,000 &  2.0  $\times\ 10^{-6}$\\
S Per    & 11.39 & 0.66 & 1.226 & 20 & 1.90 &   40.59 & 3.67 &  81\,000 &  86\,000 &  6.8  $\times\ 10^{-6}$\\ 
V441 Per & 11.40 & 0.66 & 0.156 & 16 & 1.90 &    3.54 & 1.69 &  66\,000 &  50\,000 &  4.2  $\times\ 10^{-7}$\\  
YZ Per   & 11.40 & 0.66 & 0.291 & 16 & 1.90 &    5.28 & 1.86 &  48\,000 &  55\,000 &  6.5  $\times\ 10^{-7}$\\
W Per    & 11.40 & 0.66 & 0.495 & 16 & 1.90 &   14.87 & 2.55 &  54\,000 &  56\,000 &  2.1  $\times\ 10^{-6}$\\ 
BD+57647 & 11.40 & 0.66 & 0.351 & 14 & 1.90 &    6.47 & 2.24 &  80\,000 &  37\,000 &  9.3  $\times\ 10^{-7}$\\ 
NO Aur   & 10.70 & 0.47 & 0.146 & 18 & 1.38 &    5.12 & 1.62 &  67\,000 &  73\,000 &  2.9  $\times\ 10^{-7}$\\ 
$\alpha$ Ori&  5.57 & 0.18 & 0.177 & 15 & 0.130 &  299.00 & 1.64 &  ...&  56\,000 &  1.5  $\times\ 10^{-7}$\\  
TV Gem   & 10.70 & 0.66 & 0.294 & 19 & 1.38 &    6.06 & 1.69 & 100\,000 &  84\,000 &  3.5  $\times\ 10^{-7}$\\ 
BU Gem   & 10.70 & 0.66 & 0.248 & 19 & 1.38 &   10.50 & 1.62 &  83\,000 &  86\,000 &  5.9  $\times\ 10^{-7}$\\ 
V384 Pup & 13.00 & 0.57 & 0.260 & 18 & 4.00 &    2.76 & 1.84 &  37\,000 &  75\,000 &  1.4  $\times\ 10^{-6}$\\ 
CK Car   & 11.70 & 0.55 & 0.531 & 22 & 2.20 &   13.98 & 2.11 & 161\,000 & 123\,000 &  2.1  $\times\ 10^{-6}$\\   
V602 Car & 11.60 & 0.48 & 0.596 & 22 & 2.10 &   12.40 & 2.61 & 105\,000 & 124\,000 &  1.9  $\times\ 10^{-6}$\\ 
V396 Cen & 11.60 & 0.75 & 0.201 & 23 & 2.10 &    4.98 & 1.73 & 164\,000 & 140\,000 &  6.2  $\times\ 10^{-7}$\\  
KW Sgr   & 12.40 & 0.92 & 1.218 & 27 & 3.00 &   18.39 & 2.81 & 363\,000 & 228\,000 &  5.6  $\times\ 10^{-6}$\\
NR Vul   & 11.80 & 0.94 & 0.590 & 21 & 2.30 &   12.28 & 2.26 & 224\,000 & 111\,000 &  2.2  $\times\ 10^{-6}$\\
BI Cyg   & 11.00 & 0.93 & 0.671 & 22 & 1.58 &   51.23 & 2.67 & 226\,000 & 123\,000 &  4.6  $\times\ 10^{-6}$\\  
KY Cyg   & 11.00 & 0.93 & 0.702 & 22 & 1.58 &   50.74 & 3.15 & 272\,000 & 138\,000 &  4.9  $\times\ 10^{-6}$\\ 
RW Cyg   & 10.60 & 1.22 & 0.481 & 23 & 1.32 &   60.69 & 1.96 & 144\,000 & 145\,000 &  3.2  $\times\ 10^{-6}$\\  
$\mu$ Cep&  9.70 & 0.69 & 0.361 & 20 & 0.87 &  127.00 & 1.69 & 340\,000 & 410\,000 &  1.4  $\times\ 10^{-6}$\\   
V354 Cep & 12.20 & 0.63 & 0.566 & 18 & 2.75 &    8.01 & 2.89 & 369\,000 &  76\,000 &  2.4  $\times\ 10^{-6}$\\   
V355 Cep & 12.20 & 0.63 & 0.292 & 14 & 2.75 &    3.27 & 2.38 &  94\,000 &  37\,000 &  1.0  $\times\ 10^{-6}$\\ 
PZ Cas   & 11.90 & 0.69 & 1.217 & 30 & 2.40 &   96.48 & 3.57 & 212\,000 & 193\,000 &  2.6  $\times\ 10^{-5}$\\  
TZ Cas   & 11.90 & 0.69 & 0.541 & 19 & 2.40 &    9.47 & 2.38 &  98\,000 &  83\,000 &  2.0  $\times\ 10^{-6}$\\ 
         &       &      &       &      &       &         &       &        &        &\\
EV Car   & 13.13 & 0.51 & 0.745 & 39 & 4.20 &   25.87 & 2.57 &   ...  & 675\,000 & 1.3   $\times\ 10^{-5}$\\
HS Cas   & 12.00 & 0.83 & 0.275 & 17 & 2.50 &    3.51 & 1.92 &   ...  &  59\,000 & 7.5   $\times\ 10^{-7}$\\  
XX Per   & 11.14 & 0.66 & 0.492 & 16 & 1.69 &    4.23 & 2.03 &   ...  &  50\,000 & 4.3   $\times\ 10^{-7}$\\  
KK Per   & 11.14 & 0.66 & 0.117 & 16 & 1.69 &    2.23 & 1.59 &   ...  &  50\,000 & 2.0   $\times\ 10^{-7}$\\  
AD Per   & 11.14 & 0.66 & 0.164 & 15 & 1.69 &    2.85 & 1.66 &   ...  &  42\,000 & 2.7   $\times\ 10^{-7}$\\ 
PR Per   & 11.14 & 0.66 & 0.131 & 14 & 1.69 &    2.37 & 1.55 &   ...  &  34\,000 & 2.3   $\times\ 10^{-7}$\\ 
GP Cas   & 11.40 & 0.66 & 0.200 & 15 & 1.90 &    4.45 & 2.01 &   ...  &  43\,000 & 5.9   $\times\ 10^{-7}$\\
VY CMa   & 10.28 & 0.47 & 6.959 & 47 & 1.14 & 1453.00 & 7.77 &   ...  & 295\,000 & 1.6   $\times\ 10^{-4}$\\
$\alpha$ Sco  &  6.34 & 0.10 & 0.165 & 17 & 0.185 &  115.50 & 1.73 &    ...  &  71000 & 1.2 $\times\ 10^{-7}$\\  
VX Sgr   & 10.98 & 0.52 & 2.334 & 25 & 1.57 &  262.70 & 4.14 &  ...   & 343\,000 &  2.0  $\times\ 10^{-5}$\\
Case 49  & 11.70 & 0.87 & 0.155 & 15 & 2.19 &    6.58 & 2.01 &  ...   &  42\,000 &  1.1  $\times\ 10^{-6}$\\
U Lac    & 12.20 & 0.63 & 0.685 & 23 & 2.75 &    9.04 & 2.51 &  ...   & 147\,000 &  2.3  $\times\ 10^{-6}$\\

\noalign{\smallskip}
\hline
\end{tabular}
\end{center}
\smallskip
\end{table*}


In order to compare the mass-loss rate of these RSGs with the de Jager prescription,
we estimate \Mdot\, from the infrared excess at 60\,$\mu$m  with the formulation 
of Jura and collaborators (JK90):

\smallskip
\smallskip
 
  $\Mdot_{60\mu}$$_{\rm m}$\,=\,1.7\,$\times$\,$10^{-7}$ $V_{15}$\,$D_{\rm kpc}^2$\,$f_{60}$\,
 $(\lambda_{10}/L_{4})^{1/2}$,\\

\smallskip
\smallskip

 where \Mdot$_{60\mu}$$_{\rm m}$\, is in \Msolyr,  $V_{15}$ is the wind 
expansion velocity normalized to 15 \kms, $D_{\rm kpc}$ is the distance in kpc, 
$f_{60}$ is the IRAS flux at 60\,$\mu$m in Jy (assumed to dominate the  photospheric 
flux at this wavelength), $\lambda_{10}$ is equal to  the mean wavelength
$\lambda_{\rm m}$ of the object spectral energy distribution, normalized to 10\,$\mu$m, 
 and $L_{4}$ is the luminosity in units of 10$^{4}$ \Lsol.
 This formula is valid for  a gas-to-dust mass ratio of 200, which corresponds to complete
condensation of refractory elements for solar abundances, and we adopt this ratio here.

 The wind speed in  Jura's formula is known for only a few RSGs 
(see Table 2). For all others, we choose to scale the value for this speed with
the luminosity, and we adopt $V \propto L^{0.35}$ following predictions of
dust-driven wind modelling (Habing et al.\,\,1994). For the RSGs of the solar 
neighbourhood, we find that $V \simeq 20$\,$\times$~$(L/10^{5})^{0.35}$ km~s$^{-1}$ 
roughly represents the  observed values. For the LMC RSGs considered below,
we find $V \simeq 14$\,$\times$~$(L/10^{5})^{0.35}$ km~s$^{-1}$ (see Appendix~C for details).

Concerning the luminosity  and mean wavelength, these quantities are calculated by
integrating over the spectral energy distribution. We use the UBVIJHKL  magnitudes and IRAS
fluxes and correct the magnitudes for interstellar reddening following the extinction law
of Cardelli et al.\,\,(1989). Between the central wavelengths of these spectral bands, 
interpolation is done linearly on the relation of   log($\lambda\ f_{\lambda}$) as a
 function of log($\lambda$).
The accuracy of the  derived luminosities (listed as $L_{\rm phot}$ in Table 1) depends on
several factors. Once the distance is fixed with the distance modulus, 
factors like photometric uncertainties, variability  or errors on  $E_{\rm B-V}$ 
introduce uncertainty. For example, adopting JHK magnitudes smaller by 0.2 mag 
results in larger luminosity by $\la$ 11\%, and decrease of \Mdot\ by $\la$ 7\%. 
Similarly, increasing $E_{\rm B-V}$ by 0.25~mag implies an increase in $L$  
by $\la$ 40\%, and a decrease in \Mdot\ by $\la$ 16\%.
These variations are smaller than the accuracy of the Jura's formula.

In Table 1, we have included the luminosities from LM05 (column $L_{\rm Lev}$ 
 for the  26 objects originating
from their study). Many determinations of luminosity are in very good agreement between
theirs and ours. However, there are also a few significant discrepancies, up to 
a factor of 5 for V354 Cep. Because we use the same distance and reddening, and almost 
the same photometry, the discrepancy
must originate from the method for deriving $L$, but
we could not trace exactly where the differences come from. Consequently, we used 
our luminosities for plotting $\Mdot$\ versus $L$ (Fig.~2).

 It can be seen in Fig.~2 that, if we omit four exceptions labelled 1,2,3 and 7, 
the  majority of points are clustered near the de Jager prescription line. 
For log\,$L$\,$<$\,5.2,  the de Jager line slightly overestimates the mass-loss rates (by
a factor\,$\sim$\,1.5), while for log\,$L$\,$>$\,5.2, the de Jager line passes
below the stars PZ Cas and VY CMa, but above KW Sgr and VX Sgr, and
increasing the dJ88 rate by 1.5 would not improve the situation.

The stars $\alpha$~Ori, $\alpha$~Sco and $\mu$ Cep (labelled 1, 2, 3 respectively)  
present a low \Mdot\ here, but observations of their circumstellar gas suggest that
 they are plausibly depleted in grains (see next section).  We note that our
 estimate of $\Mdot$\, for $\mu$ Cep, based
on the 60~$\mu$m flux, is in good agreement with that of de Wit et al.\,\,(2008)
derived from imaging at 24~$\mu$m, after appropriate scaling of the outflow velocity.

 There is also one star at very high luminosity: EV Car (labelled 7),  with 
$L =  6.7$\,$\times$\,$10^{5}$\,\Lsol\, and
$\Mdot\,$\,$=$\,$1.3 \times 10^{-5}$\,\Msolyr. The huge luminosity is due to the distance
modulus given by Humphreys (1978),   $(m-M) = 13.3$, corresponding to 4.2 kpc. This puts
 the star far beyond the Car OB1 association. If we had taken
the distance of this association (2.5~kpc, adopted by JK90), 
we would have found $L = 2.4 \times 10^{5}$\,\Lsol\, 
and $\Mdot = 4.6 \times 10^{-6}$ \Msolyr, and 
the location of the star in Fig.~2 would be very close to the de Jager line. 
But to our knowledge, there is no observational evidence for EV Car 
to be member of Car OB1.


\begin{figure}
\includegraphics*[width=7cm,angle=-90]{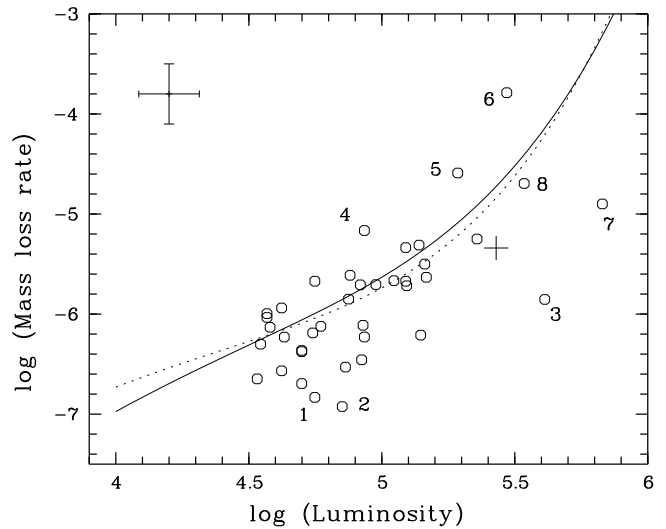}

\caption[]{Comparison of de Jager rate (plotted for $T_{\rm eff} = 3500$\ K 
and 4000\ K) with mass-loss rates of 39 RSGs listed in Table 1. The mass-loss rates 
are based on the study of dust and obtained with Jura's formula. The labels indicate
a few wellknown RSGs as follows: 1:\ $\alpha$~Ori; 2:\ $\alpha$~Sco, 3:\ $\mu$~Cep, 4:\ S Per,
 5:\ PZ Cas, 6: VY CMa, 7:\ EV Car (at 4.2 kpc); 8: VX Sgr.  The small cross represents EV Car if
a smaller distance of  2.5 kpc is adopted. At the upper left corner, error bars represent
 uncertainties of $\pm$\ 30 percent in luminosity, and a factor of $\pm$ 2 in \Mdot.}
\label{fig02}
\end{figure}

\subsection{Circumstellar gas of RSGs in the solar neighbourhood}

Only a few RSG winds have been detected through observation of the circumstellar gas.
Methods of detection include measurements of the {\ion{H}{i}} 21cm emission 
(details on observations and mass-loss determination can be found in Bowers \& Knapp~1987),
 millimetric emission of CO (Olofsson~2004; Ramstedt et al.~2008), submillimeter 
emission of {\ion{C}{i}} (Keene et al.~1993; Huggins et al.~1994), 
 emission of scattered resonance lines of \NaI\ and \KI\ (Mauron~1990; Guilain \& Mauron 1996;
 Gustafsson et al.~1997), and 
observation of the \HII\  region around Antares (Kudritzki \& Reimers 1978; Reimers et al.~2008).

For eight stars, we have collected in Table~2 the mass-loss rates obtained with 
these methods, giving detailed reference for adopted parameters ($D$ and  $E_{\rm B-V}$) 
used to derive luminosities. When $\Mdot$\, was obtained in original papers with a different 
distance, we assumed $\Mdot$\,$\propto$\,$D^{2}$. The expansion velocities $V$ are from
Huggins et al.\,\,(1994) for $\alpha$ Ori, Woodhams (1993) for S Per and  PZ Cas, and
from Kemper et al.\,\,(2003) for  VY~CMa, VX~Sgr, NML~Cyg. 
Finally, for $\mu$ Cep, the situation is unclear (for details 
see Bernat 1981, the discussion in Le Borgne \& Mauron 1989, and Kemper et al.\,\,2003)
 and  we have adopted 20 \kms. 

Several remarks have to be made on individual objects.
For the most observed star ($\alpha$ Ori), there is good agreement among the rates based on 
\HI, \CI\ and \KI\ (Bowers \& Knapp 1987; Huggins et al.\,\,1994; Glassgold \& Huggins 1986; 
Plez \& Lambert 2002). When the distance given by Hipparcos is adopted 
(parallax 7.63\,$\pm$\,1.64 mas, giving 130\,pc\,$\pm$\,30\,pc),  one obtains
$\Mdot$~$\approx$~1~to~2\,$ \times$\,$  10^{-6}$\,\Msolyr. The distance proposed 
by Harper et al.\,\,(2008) is  197 $\pm$ 45 pc and would imply 
$\Mdot \sim 3 \times  10^{-6}$\,\Msolyr. For this star, the interstellar 
extinction $A_{\rm V}$ is not very well determined, being  between 0.3 and 0.8 mag 
according to Lambert et al.\,\,(1984). We have chosen the average of these values 
implying  $ L = 56000$\,\Lsol, but the main uncertainty comes from the error on 
parallax, and $L$ is known within $\pm$\,50 percent.

In the case of VX Sgr, the distance has recently been measured by Chen et al.\,\,(2007) with
SiO maser proper motions ($D = 1.57 \pm 0.27$ kpc), and this distance means that the 
 star belongs to the Sgr OB1 association. The average interstellar
 extinction to the OB stars of this association is
 $E_{\rm B-V} = 0.5$ (from data of Humphreys 1978), and the  luminosity 
is 3.4\,$\times$\,10$^5$ \Lsol.

In the case of VY CMa, a distance of 1.14\,$\pm$\,0.10\,kpc 
has been obtained by Choi et al.\,\,(2008)
through astrometric monitoring of H$_{2}$O masers. This distance is smaller than often used.
For example, Decin et al.\,\,(2006) have adopted 1.50~kpc for their detailed study of 
circumstellar CO lines. The ejection rate  obtained by Decin et al.\,\,is highly variable. 
The episode with the highest mass-loss rate
($\sim 3 \times 10^{-4}$\,\Msolyr) has a duration of only 100~yr.  The average mass-loss rate is
about 7\,$\times$\,10$^{-5}$\,\Msolyr. Scaled to the distance of 1.14\,kpc, the average rate becomes
4\,$\times$\,10$^{-5}$\,\Msolyr. The new smaller distance means that the star is less luminous than
previously believed. We find $L = 2.9 \times 10^{5}$\,\Lsol. 
 Because the star is extremely red (the flux averaged wavelength is around 8$\mu$m, compared 
to $\sim 2 \mu$m\ for the majority of objects),
this luminosity does not depend much on the choice of the extinction:
 for $E_{\rm B-V}$ in the range  0 to 1, the luminosity is between  $L = 2.7$ and
$3.5 \times 10^{5}$\,\Lsol.

The points representing these eight stars with measured gas mass-loss rates as a 
function of luminosity are shown in Fig.~3. The error bars    correspond to $\pm 30$  
percent in luminosity and a factor of $\pm$\,2 of uncertainty on the mass-loss rate.  
The actual uncertainties on $\Mdot$\ are probably larger than that but difficult to quantify. 
The points are  not very far from the de Jager line. Betelgeuse and Antares agree for 
$\Mdot \approx 10^{-6}$\,\Msolyr\ at a luminosity of $\sim$~65000\,\Lsol.  S\,Per is a factor of 4  
above the line  and $\mu$\,Cep is a factor of 10 below the line.  It can be seen that on average 
the de Jager prescription agrees reasonably well with measurements of the gas mass-loss rates.

        \begin{table*}[!ht]
        \caption[]{Characteristics of RSGs for which a mass-loss rate  
has been measured by observing the circumstellar gas.} 
         \label{table2}

        \begin{center}   
        \begin{tabular}{lllcrcrrrcccc}
        \hline
        \hline
        \noalign{\smallskip}

   Name & $D$ & $E_{\rm B-V}$ & V  & $F_{60}$  &  $\lambda_{\rm m}$ & $L_{\rm phot}$ & \Mdot$_{60\mu}$$_{\rm m}$ & \Mdot$_{\rm gas}$    &  Method & \multicolumn{3}{c}{References}\\
        & [kpc]& [mag]    & [km s$^{-1}$] & Jy & [$\mu$m] & [\Lsol]      & \multicolumn{2}{c}{ [10$^{-6}$\,\Msolyr]} &          &         &      & \\
   (1)  &  (2) &  (3)     &  (4)          & (5)&  (6)     & (7)          & (8)      & (9)       & (10)     & (11)    & (12) & (13)\\ 

       \noalign{\smallskip}
        \hline
        \noalign{\smallskip}

$\alpha$ Ori  & 0.130 & 0.18  & 15 & 299  & 1.64  &  56\,000  & 0.15 & 1.5   & \HI\, \CI\, \KI\,    & 1 & 2 & 3\\  
$\alpha$ Sco  & 0.185 & 0.10  & 17 & 115  & 1.73  &  71\,000  & 0.12 & 1.0   & \HII\,reg.    & 1 & 4 & 5\\
S Per         & 1.90  & 0.66  & 20 & 41   & 3.67  &  86\,000  & 6.8  & 7.5   & CO          & 6 & 6 & 7\\ 
PZ Cas        & 2.40  & 0.69  & 30 & 96   & 3.57  & 193\,000  & 26  & 8.3     & CO            & 8 & 8 & 7\\
VY CMa        & 1.14  & 0.47  & 47 & 1453 & 7.77  & 295\,000  &160   & 40    & CO          & 9 & 10 & 11\\
NML Cyg       & 1.74  & 1.20  & 34 & 1030 & 9.80  & 320\,000  &210   & 140   & CO          & 12 & 13& 14\\
VX Sgr        & 1.57  & 0.52  & 25 &  263 & 4.14  & 340\,000  &20    & 11    & CO          & 15  & 16 & 17\\
$\mu$\,Cep   & 0.87  & 0.69  & 20 &  127 & 1.69  & 410\,000  &1.4   & 5.0   & \KI\, \NaI      & 6 & 6 & 18\\

  \noalign{\smallskip}
   \hline
\end{tabular}
\end{center}
 Notes: Col.~(11) = reference for distance. Col.~(12) = reference for $E_{\rm B-V}$.
 Col.~(13) = reference for mass-loss rate $\Mdot_{\rm gas}$. The IRAS data of NML Cyg were taken
from Schuster et al.\,\,(2009).

\smallskip

References: (1) Hipparcos parallax; (2) Lambert et al.\,\,1984; (3) Huggins et al.\,\,1994;
            (4) assumed; (5) Reimers et al.\,\,2008; (6) LM05; (7) Woodhams 1993; (8) JB00;
            (9) Choi et al.\,\,2008; (10) assuming $E_{\rm B-V} = 0.47$ per kpc;
            (11) Decin et al.\,\,2006 \& see text; (12) Massey \& Thomson 1991; 
            (13)  Bl\"{o}cker et al.\,\,2001; (14) Knapp et al.\,\,1982; (15) Chen et al.\,\,2007;
            (16) Humphreys 1978; (17) Knapp et al.\,\,1989; (18) Guilain \& Mauron 1996. 

\end{table*}

\begin{figure}
\includegraphics*[width=7cm,angle=-90]{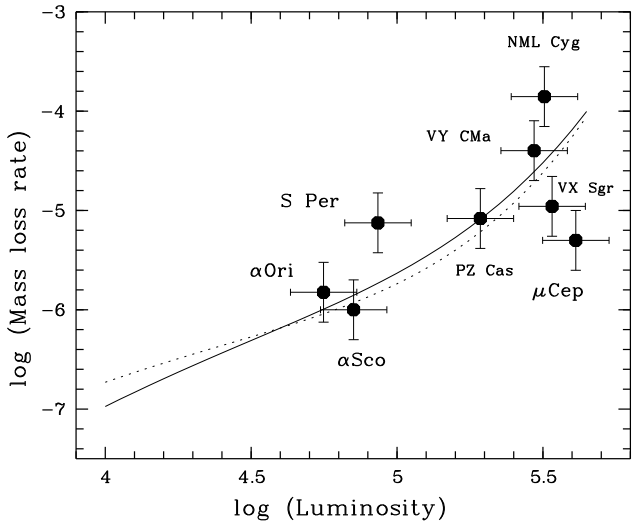}
\caption[]{ Mass-loss rates (from observations of the gas) 
$versus$ luminosity for  the eight RSGs
 of Table 2. The lines are the de Jager rates for 4000K (solid line) and 3500K (dotted line).}
\end{figure}

 Since $\Mdot_{60\mu}$$_{\rm m}$\, was used in the previous section, we can  
compare it,  for these stars, with  the   mass-loss rates derived from observation of
the gas $\Mdot_{\rm gas}$. This is done in Fig.~4 and Table~2.  Figure~4 shows that the
agreement between the two rates is satisfactory for some stars (S Per, NML Cyg, VX Sgr), 
but there are also differences of a factor of $\sim$~4 for VY CMa and $\mu$~Cep.  
These differences may arise from the fact that 
different regions of the envelope are probed when comparing different techniques. 
For $\alpha$~Ori and $\alpha$ Sco, $\Mdot_{60\mu}$$_{\rm m}$ is too low by
a factor of $\sim$10, plausibly due to incomplete dust formation. 
So, there is a trend in Fig.~4 that  for low \Mdot, grain condensation may be incomplete, but
obviously many additional sources should be examined with independent estimates of the 
dust mass-loss rate
and the gas mass-loss rate to confirm this trend.

\begin{figure}
\includegraphics*[width=7cm,angle=-90]{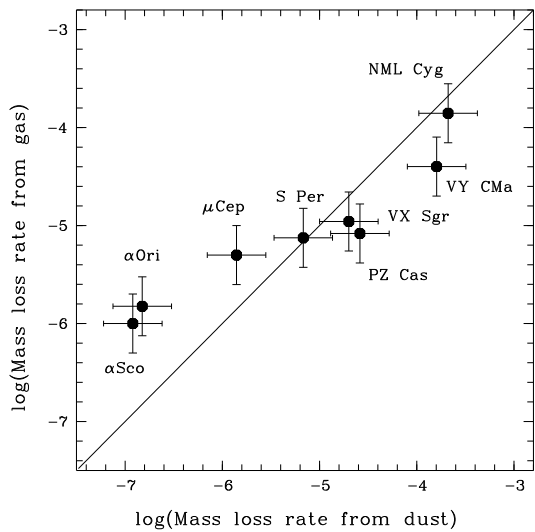}
\caption[]{ Mass-loss rates from observation of the gas $versus$ mass-loss rate  
from observation of the dust (the 60\,$\mu$m emission) 
 for  the eight RSGs  of Table 2.}
\end{figure}

 \subsection{Infrared measurements in the Large Magellanic Cloud }

The  mass-loss rate of RSGs in the Large Magellanic Cloud (LMC)  has been
investigated in several works, with various data and various methods.
All works are based on the circumstellar dust emission.
Reid et al.\,\,(1990) used IRAS fluxes and the Jura's formula. van Loon et al.\,\,(2005)
exploited MSX, ISO and IRAS fluxes (their Table 3) and analysed the spectral
energy distribution with the dust radiative transfer code DUSTY, 
which provides \Mdot. More recently, Groenewegen et al.\,\,(2009) exploited 
Spitzer results in addition to all previously obtained data.
In particular,  Spitzer data include photometry from 3.6 to 24~$\mu$m  and 
spectroscopy  from 5.2 to 37~$\mu$m. Groenewegen et al.\,\,(2009) fit the data
also with dust radiative transfer models.(The work of Bonanos et al.\,\,(2010)
also treats mass loss of RSGs in the Magellanic Clouds and
is considered in the Discussion.)

The mass-loss rates given in the works cited above cannot be compared directly,
because their assumed parameters, such as the expansion velocity of the wind or
the gas-to-dust mass ratio  are different. We assume here that the LMC RSGs have
a metallicity  lower than that of the Sun so that complete condensation of
refractory metals leads to a gas-to-dust mass ratio $\psi$\,$=$\,$500$. 
  We also assume that all RSGs have
a wind velocity $v(L)$ depending on the luminosity $L$ as found in Appendix~C 
i.e. $v(L) = 14 (L/10^5)^{0.35}$\,\kms. 
 Consequently, we have multiplied the
 \Mdot\, of Reid et al.\,\,(1990) by 2.5\,$\times$\,$v(L)/15$ (because the Jura's 
formula used by Reid et al.\,\,supposes $\psi = 200$ and a wind speed of 15~\kms).
We have multiplied the \Mdot\, of van Loon et al.\,\,(2005) by $v(L)/v'$ where $v'$ is their
published wind velocity provided by DUSTY. Finally,
 we have multiplied by 2.5 $\times$ $v(L)/10$ the \Mdot\, of Groenewegen et al.\,\,(2009)
because they adopt $\psi = 200$ and  a wind speed of 10 km\,s$^{-1}$.

 The results are in Table 3, 4 and 5. As far as possible, we did not include AGB stars
in these tables. More precisely, we followed Groenewegen et al.\,\,(2009) by considering
that all objects with $L \ga 1.2 \times 10^5$ \Lsol\ are genuine RSGs, and that 
objects with $L$ between
$\sim 5 \times 10^{4}$ and 1.2 $\times$ 10$^{5}$ \Lsol\ can be considered as RSGs only
if their amplitude of variation in the optical $I$ band was less than 0.45 mag. This selection
implied that 7 objects in the Groenewegen sample were kept 
(marked with ``s'' after their names in the first column of Table 5) and 9 were rejected.

\begin{table}[!ht]
\caption[]{Red supergiants of Reid et al.\,\,(1990). The columns provide
the object name, J2000 coordinates as for 2MASS names, luminosities in \Lsol,
mass-loss rates in units of 10$^{-6}$ \Msolyr, and the ratios $r$ with 
mass-loss rates  from Groenewegen et al.\,\,(2000) for common objects.}
\begin{center}
\begin{tabular}{lrrrrr}
\hline \hline
 Name   &  $\alpha, \delta$~(J2000)\,\,\,\,\,\,\,\,\, &  $L$\,\,\,      & $\Mdot$ & $r$\\
\hline
\noalign{\smallskip}
\object{HV 12501} & 045541.83$-$692624.3 & 132\,000 & 157& ..\\
\object{HV 12420} & 045731.52$-$700859.2 & 110\,000 & 79 & ..\\
\object{HV 2255}  & 045743.32$-$700850.3 & 251\,000 & 196& 37\\
\object{HV 888}   & 050414.14$-$671614.4 & 328\,000 & 134& 37\\
\object{HV 894}   & 050533.50$-$703347.0 & 106\,000 & 86 & ..\\
\object{HV 2360}  & 051246.37$-$671937.9 & 161\,000 & 135& ..\\
\object{HV 916}   & 051449.74$-$672719.8 & 148\,000 &  96& 37\\
\object{HV 2532}  & 052627.39$-$691055.9 & 113\,000 &  49& ..\\
\object{HV 963}   & 052734.35$-$665330.0 & 100\,000 &  56& 31\\
\object{HV 5854}  & 052815.41$-$665852.7 &  97\,000 &  23& ..\\
\object{HV 2586}  & 052934.43$-$665528.3 &  71\,000 &  50& ..\\
\object{HV 996}   & 053235.61$-$675508.9 & 127\,000 & 172& 26\\
\object{HV 12437} & 053307.61$-$664805.7 &  82\,000 &  65& ..\\
\object{HV 2700}  & 053518.92$-$670219.6 &  89\,000 &  42& ..\\
\object{HV 1004}  & 053625.46$-$665538.4 & 141\,000 & 110& ..\\
\noalign{\smallskip}
\hline
\end{tabular}
\end{center}
\end{table}

\begin{table}[!ht]
\caption[]{Red supergiants of van Loon et al.\,\,(2005). Same columns as in Table 3.}
\begin{center}
\begin{tabular}{lrrrr}
\hline \hline
  Name     &  $\alpha, \delta$~(J2000)\,\,\,\,\,\,\,\,\, & $L$\,\,\,\,   & $\Mdot$ & $r$\\
\hline
\noalign{\smallskip}
\object{WOH-G64}    & 045510.48$-$682029.8 &  490\,000 & 1200 & 1.6\\
\object{HV 12501}   & 045554.83$-$692624.3 &  145\,000 & 15 & ..\\
\object{HV 888}     & 050414.14$-$671614.4 &  269\,000 & 37 & 10\\
\object{HV 2360}    & 051246.37$-$671937.9 &  129\,000 & 30 & ..\\
\object{HV 916}     & 051449.74$-$672719.8 &  148\,000 & 35 & 13\\
\object{HV 2561}    & 052828.86$-$680707.9 &  179\,000 & 49 & 16\\
\object{HV  5870}   & 052903.48$-$690646.2 &   95\,000 & 23 & 8\\
\object{SP77-46-44} & 052942.21$-$685717.4 &  295\,000 & 52 & 37\\
\object{HV 986}     & 053109.28$-$672554.9 &  132\,000 & 20 & ..\\
\object{HV 996}     & 053235.61$-$675508.9 &  117\,000 & 63 & 10\\
\noalign{\smallskip}
\hline
\end{tabular}
\end{center}
\end{table}

\begin{table}[!ht]
\caption[]{Red supergiants of Groenewegen et al.\,\,(2009), with
names, J2000 coordinates, luminosities and mass-loss rates in units of 
10$^{-6}$ \Msolyr.}
\begin{center}
\begin{tabular}{lrrr}
\hline
\hline
 Name &  $\alpha, \delta$~(J2000)\,\,\,\,\,\,\,\,\,  & $L$\,\,\,\,  & $\Mdot$\\
\hline
\noalign{\smallskip}
\object{HV 2236}       & 044922.46$-$692434.4 &   120\,000 & 2.8\\
\object{MSX LMC 1189}  & 045503.07$-$692912.7 &   139\,000 & 6.7\\
\object{WOH-G64}       & 045510.48$-$682029.8 &   440\,000 & 764\\
\object{MSX LMC 1204}  & 045516.05$-$691912.1 &   204\,000 & 0.81\\
\object{MSX LMC 1318} s& 045533.54$-$692459.3 &    61\,000 & 1.3\\
\object{HV 2255}       & 045743.31$-$700850.3 &   143\,000 & 5.3\\
\object{HV 888}        & 050414.13$-$671614.3 &   337\,000 & 3.6\\
\object{HV 916}        & 051449.72$-$672719.7 &   250\,000 & 2.6\\
\object{WOH-S264}      & 052419.32$-$693849.3 &   215\,000 & 4.5\\
\object{MSX LMC 549}   & 052611.35$-$661211.1 &   132\,000 & 2.0\\
\object{MSX LMC 575}   & 052622.18$-$662128.5 &   163\,000 & 0.20\\
\object{MSX LMC 589}   & 052634.80$-$685140.0 &   269\,000 & 1.7\\
\object{HV 963}       s& 052734.35$-$665330.0 &   115\,000 & 1.8\\
\object{HV 2561}       & 052828.86$-$680707.8 &   139\,000 & 3.0\\
\object{HV 5870}      s& 052903.48$-$690646.2 &    79\,000 & 2.8\\
\object{W60 A27}       & 052942.21$-$685717.3 &   280\,000 & 1.4\\
\object{MSX LMC 810}   & 053020.67$-$665301.8 &   197\,000 & 1.1\\
\object{MSX LMC 587}   & 053104.18$-$691903.0 &   218\,000 & 2.3\\
\object{W60 D22}       & 053110.64$-$663531.6 &   163\,000 & 0.66\\
\object{MSX LMC 839}   & 053136.81$-$663007.6 &   255\,000 & 2.2\\
\object{HV 996}        & 053235.61$-$675508.9 &   132\,000 & 6.6\\
\object{MSX LMC 791}  s& 053524.52$-$690403.4 &    84\,000 & 0.89\\
\object{MSX LMC 870}   & 053528.32$-$665602.4 &   218\,000 & 2.3\\
\object{MSX LMC 891}   & 053555.22$-$690959.4 &   265\,000 & 3.1\\
\object{W60 A72}      s& 053833.97$-$692031.7 &   115\,000 & 84\\
\object{W60 A78}       & 053932.33$-$693450.0 &   223\,000 & 0.79\\
\object{MSX LMC 897}   & 054043.75$-$692158.1 &   245\,000 & 1.0\\
\object{MSX MC 939}   s& 054048.50$-$693336.1 &   109\,000 & 4.4\\
\object{HV 1017}      s& 054059.20$-$691836.2 &   102\,000 & 4.3\\
\object{HV 2834}       & 054413.73$-$661644.5 &   215\,000 & 1.1\\
\noalign{\smallskip}
\hline
\end{tabular}
\end{center}
\end{table}

 Tables 3, 4 and 5 include names, J2000 coordinates, luminosities, 
mass-loss rates (scaled as explained above and in units of 10$^{-6}$ \Msolyr). 
The last column in Table 3 and 4 are the ratio of the \Mdot\, from
Reid et al.\,\,(or from van Loon et al.\,) and the \Mdot\, from Groenewegen et al.\,\,(2009) 
for objects belonging to both samples. 

Figure~5 shows these scaled \Mdot\, as a function of $L$ from the three works 
mentioned above. We have also shown the de Jager rate for 4000 and 3500 K exactly as
for previous plots. It can be seen immediately that there is a  difference between the 
recent rates from Groenewegen sample  on one hand (the median rate is 
2.3 $\times$  10$^{-6}$ \Msolyr, 30 RSGs represented by circles),
and, on the other hand, the Reid sample (median 9.0 $\times$ 10$^{-5}$\, \Msolyr; 15 crosses) 
or the van Loon sample (median 3.6  $\times$ 10$^{-5}$\, \Msolyr; 10 small squares). 
One can also consider the objects in common (see Table 3 and 4): these objects 
indicate that the Reid mass-loss rates are $\sim$ 35 times larger than those of 
Groenewegen, and that the van Loon mass-loss rates are $\sim$ 10 times larger than 
those of Groenewegen. One notes however that, for WOH-G64, the \Mdot\, from van Loon 
and Groenewegen agree reasonably  well (within a factor 1.6). In contrast, 
for SP77-46-44 (alias W60 A27), they disagree by a factor 37.
  
 There are two  assumptions in Reid et al.\,\,(1990) which can explain (at least in part)
why their rates are so high. First, there is probably an  overestimate 
of the flux averaged wavelength: Reid et al.\,\,(1990) take $\lambda_{10}=1$, which
 is not reached even by NML Cyg (see Table 2).  For the optical RSGs making the Reid sample, 
$\lambda_{10}$ is likely close to 0.2 only. Secondly, Reid et al.\,\,(1990) 
overestimate  the $f_{60}$/ $f_{25}$ IRAS flux ratio used to
obtain $f_{60}$.  They take 0.50 instead of $\sim$ 0.20. A ratio of 0.23 $\pm  0.03$  
is reached only by dusty Galactic RSGs (VX Sgr: 0.19, VY CMa: 0.22, PZ Cas: 0.24, NML Cyg: 0.26).
So, the Reid et al.\,\,mass-loss rates have to be decreased by a factor $\sim 12$. 

Concerning the van Loon rates, the reasons why they are generally
larger than those of Groenewegen are not clear. With Spitzer data,
small infrared excesses become measurable and are more accurate, while 
this was not the case with MSX, ISO or IRAS 
data available for the van Loon et al.\,\,work. Therefore, there may be a bias in favour
of large \Mdot\ in the van Loon sample compared to the Groenewegen sample. 
Another possibility is that the higher spatial resolution of Spitzer allows
to  avoid source confusion and contaminating sources. This may be in particular 
the case for SP77-46-44. For this source, the IRAS fluxes at 12 and 25 $\mu$m are
0.26 and 0.18 Jy, while Spitzer photometry (from Bonanos et al.\,\,2009) indicates
0.19 Jy at 12 $\mu$m  (through interpolation from the 8~$\mu$m and 24~$\mu$m  fluxes) 
and 0.11 Jy at 25~$\mu$m. So these fainter Spitzer fluxes suggest a smaller infrared
excess and may go some way to explain the large discrepancy for this object.

 Finally, if we consider the rates by Groenewegen et al.\,\,(2009) and their positions
with respect to the de Jager lines (see Fig.~5), one can note a rather good agreement
for 12 objects with luminosities lower than 1.6 $\times$ 10$^{5}$~\Lsol (with the exception of 
W60-A72 at $L =  115000$~\Lsol, $\Mdot = 8.4 \times 10^{-5}$ \Msolyr; in Fig.~5, this object
is the circle located in the middle of Reid's plus signs). But at larger luminosities,
the Groenewegen rates are below the dJ88 lines by an average factor $\sim 8$  (13 objects),
with the exception of WOH-G64 where the de Jager rate is $\sim$~10 times lower than observed. 
It is also worthy to note that at high luminosities, two objects from the van
Loon sample are quite close to the dJ88 line.

\begin{figure}
\includegraphics*[width=7cm,angle=-90]{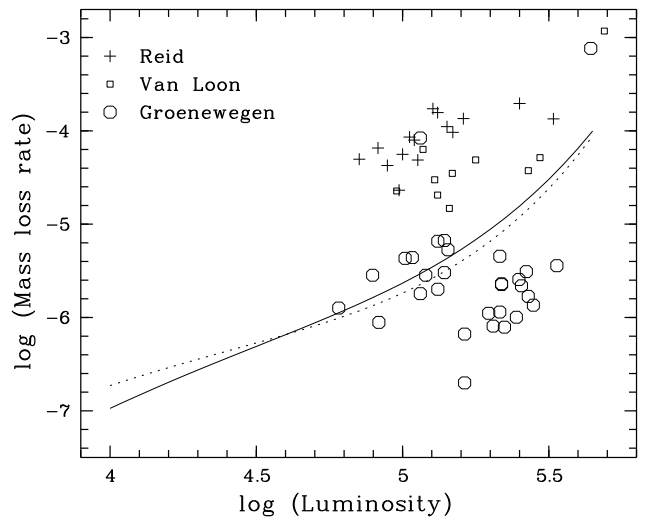}
\caption[]{ Mass-loss rates plotted as a function of luminosity for the LMC RSGs. Three
samples are plotted: Reid et al.\,\,(1990) (15 small crosses),  van Loon et al.\,\,(2005) 
(10 small squares), and  Groenewegen et al.\,\,(2009) (30 circles). 
 The de Jager rates are represented as in previous figures by
 solid and  dotted lines for 4000\,K and 3500\,K, respectively.}
\end{figure}


\section{Discussion}

The goal of this paper is to see whether the de Jager prescription
was still applicable in view of  new results obtained since 1988 and concerning
the RSGs and their winds. One important fact is that  for only
eight RSGs in the solar neighbourhood is the direct detection of the circumstellar
gas  available. In all other cases, only dust can be seen.
 We find
that for these 8 RSGs, the mass-loss rate estimated from  observations of the gas is 
in most cases within a factor of 4  of the  de Jager  prescription for
 $L < 2 \times 10^{5}$\,\Lsol (Fig.~2). For higher luminosity, it is a factor of 10.
 These deviations have to be considered in light of the following
characteristics of the RSG winds. The winds are changing with time 
(timescale $\sim$ 10$^{2}$\,-\,10$^{4}$ years),
 and  the rate of mass loss and the expansion velocity vary. This is proved
by high resolution spectroscopy of the circumstellar absorption lines which
display discrete velocity components (on $\mu$ Cep, Bernat 1981), by direct imagery
($\alpha$ Ori, Plez and Lambert 2002) or by study of the CO millimetric lines (VY CMa,
Decin et al.\,\,2006). Also, the winds  are often not spherically symmetrical, 
as revealed by some CO line profiles (PZ Cas, Woodhams 1993) or
direct imagery (HST imaging of NML Cyg,  Schuster et al.\,\,2006). 
Given these properties, the measurements of \Mdot\, with often limited 
data and simple modelling are  necessarily poor in accuracy and do not reflect 
an average mass-loss rate. Therefore, the agreement with the de Jager prescription  
seems  acceptable.

For other, much more numerous RSGs, the gas is not detectable and only dust can be seen
 through its infrared radiation. When the Jura's formula is adopted for
 deriving \Mdot,
the resulting $\Mdot$\, are also in agreement with de Jager lines, except for some
rare stars like $\mu$ Cep which has a very low dust content.
 In general, the agreement is within a factor of 4
(Fig.~2).  Part of this deviation may be due to a natural scatter of 
$\Mdot$\, at a given luminosity (see the characteristics of winds mentioned above) 
and another part may be due to the use of the Jura's formula.

\subsection{Verification of the Jura's formula}

  In order to check the Jura's formula,
 we have considered the work by Verhoelst et al.\,\,(2009) who
exploit ISO-SWS spectra (in addition to the full spectral energy distribution)
 and use state-of-the-art dust modelling.
In particular, they  include several types of grains such as 
 melilite, olivine, alumine, carbon and FeO. Together with results
concerning the dust condensation sequence in RSGs, they derive 
mass-loss rates. We took into account their assumptions on distance, expansion velocity and
 gas-to-dust mass ratio and we applied appropriate scaling, so that
 we can compare their results to ours ($\Mdot_{60\mu \rm{m}}$ in Table~2). We find that
their $\Mdot$  are 0.57, 0.88, 1.20 and 1.9 times greater than ours, for
$\alpha$ Ori, S Per, PZ Cas and $\mu$ Cep, respectively. In conclusion, 
the agreement with Jura's formula is at most a factor of 2, which is very good, 
and there is no particular reason to have doubts on this formula.

\subsection{Comparison with other mass-loss prescriptions} 

 In Sect.~2, we mentioned in Fig.~1 a few mass-loss prescriptions. After our analysis of
previous sections, we can reconsider this point. Concerning the RSGs in the solar 
vicinity, all observed mass-loss rates (from dust or gas observations) are well below 
the  rate by Salasnich et al.\,\,(1999). For example, $\alpha$\,\,Ori and $\alpha$\,\,Sco 
are located  at $L \approx 60\,000\,$\Lsol, and their $\Mdot$ are 
$\sim$ 1.5\,$\times$\,$10^{-6}$\Msolyr. The Salasnich rate would give
$3.4 \times 10^{-5}$\Msolyr, 20 times larger. Because they  are based on  mass-loss 
rates of Reid et al.\,\,which we have shown are overestimated,
the Salasnich rate and the Feast rate are too large. The Reimers rate is as good as the
dJ88 rate for  objects with $L \la 2 \times 10^5$\,\Lsol, but this is not the case for
higher luminosity objects like NML Cyg or VY CMa.

\subsection{Comparison with the van Loon formalism}

It is  appropriate to consider here the mass-loss formulation of
 van Loon et al.\,\,(2005). These authors derived a prescription for 
mass-loss rates of AGB stars and red supergiants. It is based on the 
study of evolved stars in the LMC, but  is also valid for Galactic AGB stars 
and RSGs.  An important feature of this prescription is that it is 
applicable only to {\it dusty} stars. Unfortunately, the dusty character of a 
circumstellar envelope  is not possible to be derived from the basic 
stellar parameters, i.e. $L$, $M$ and $T_{\rm eff}$.
Nevertheless, one can compare the results of this prescription to the de Jager rate.
 van Loon et al.\,\,(2005) derived  effective temperatures
from spectral classifications. Luminosities and  mass-loss rates were
determined by modelling the spectral energy distribution. The van Loon prescription is:\\

$\Mdot = 2.50 \times 10^{-5}$   $(L/10^{5})^{1.05}$    $(T/3500)^{-6.3}$\,.\\

This prescription is shown in Fig.~6 along with the de Jager lines. Both
are drawn for  $T_{\rm eff}$ of 4000, 3500 and 3000~K. van Loon et al.\,\,(2005) have
found that their recipe overestimates mass-loss rates for Galactic ``optical'' RSGs.
We can see that clearly in Fig.\,6, where the van Loon lines are well above 
the de Jager lines on which we found that most of the Galactic RSG are 
located (Fig.~2 and Fig.~3).  However,  the de Jager rate and
the van Loon rate fairly agree with each other for the highest luminosities.
Also, it is worth noting that the van Loon et al.\,\,prescription offers a significant
spread of $\Mdot$\, for a given luminosity. The spread allowed by the deJ88
prescription is clearly very small.

We have collected in Table~6 the mass-loss rates for the eight stars possessing 
an $\Mdot_{\rm gas}$ based
on observation of the gas (from Table 2). In this Table, the  $T_{\rm eff}$ values are taken 
from Lambert et al.\,\,(1984) for $\alpha$ Ori, from LM05 for S Per, PZ Cas, and $\mu$ Cep, 
from van Loon et al.\,\,(2005) for VY CMa, NML Cyg and VX Sgr, and from 
Kudritzki \& Reimers\,(1978) for $\alpha$ Sco. Column 5  of this Table gives
the de Jager rates, and column 6 gives  the van Loon rates (in column 6, the numbers
in parenthesis concern ``optical'' RSGs and   are just given for information). The ratios between
van Loon rates and the observed rates  are 2.8, 5.1, 2.2, 0.85 and 7.2  for
S Per, PZ Cas, VY CMa, NML Cyg, and VX Sgr, respectively. By considering 
the median of these numbers, the van Loon rates are $\sim$ 2.8 times too large on average. 
For comparison, the ratios between de Jager rates and observed rates for the
same stars  are  0.20, 0.78, 0.45, 0.15 and 2.9. The median is 0.45 so that, the de Jager rates 
are $\sim$ 2.2 times too small on average. In conclusion, the de Jager  and van Loon rates
are almost equally accurate (or inaccurate), but one (de Jager) underestimates and the other
(van Loon) overestimates the observed rates. The de Jager formalism is closer 
to the observed rates for non-dusty RSGs, although
it is poor in the case of $\mu$ Cep (off by an order of magnitude).

\begin{figure}
\includegraphics*[width=7cm,angle=-90]{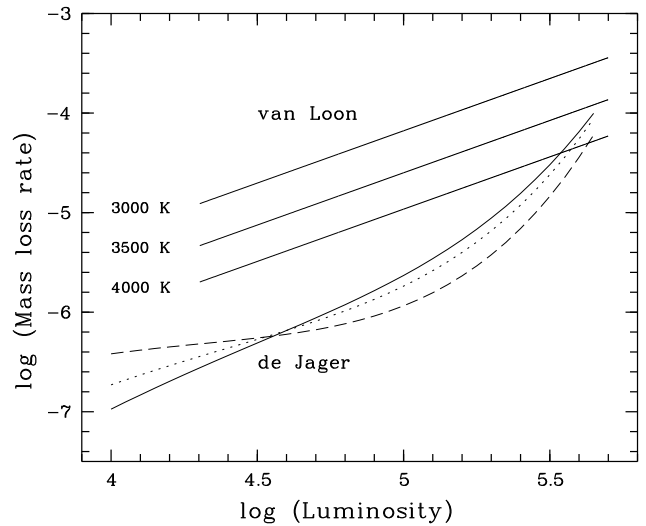}
\caption[]{ Mass-loss rate prescription of van Loon et al.\,\,(2005) and de Jager 
 (1988) with $\Mdot$ plotted as a function of luminosity and for three values of $T_{\rm eff}$.
 The de Jager lines are solid, dotted and dashed for 4000, 3500 and 3000\,K respectively. }
\end{figure}

  \begin{table}[!ht]
        \caption[]{ Red supergiants with an observed mass-loss rate derived from 
	observations of the gas ($\Mdot_{\rm gas}$).
Column 5 is the  de Jager rate, and column 6 is the rate with the van Loon recipe, which is 
applicable only for dusty stars (rates in parentheses should not be considered). Mass-loss rates
are given in units of 10$^{-6}$\,\Msolyr. } 
         \label{table3}

        \begin{center}   
        \begin{tabular}{lrrrrr}
        \hline
        \hline
        \noalign{\smallskip}
 
   Name & $L$\,\,\,\, & $T_{\rm eff}$  &  $\Mdot_{\rm gas}$   &  $\Mdot$(deJ) & $\Mdot$(VL)\\
   (1)  & (2)\,\,\,\, &  (3)\,\,\,           &   (4)\,\,\,              &   (5)\,\,\,  &  (6)\,\,\,\\  
       \noalign{\smallskip}
        \hline
        \noalign{\smallskip}

$\alpha$ Ori & 56000  & 3800 &  1.5  & 1.0   & (8) \\
$\alpha$ Sco & 71000  & 3550 &  1.0  & 1.2   & (16)\\
S Per        & 86000  & 3500 &  7.5  & 1.5   & 21 \\
PZ Cas       & 193000 & 3600 &  8.3  & 6.5   & 42 \\
VY CMa       & 295000 & 3435 & 40.0  & 18    & 88 \\
NML Cyg      & 320000 & 3310 & 140.0 & 21    & 120\\
VX Sgr       & 340000 & 3575 &  11.0 & 32    &  79\\
$\mu$ Cep    & 410000 & 3700 &   5.0 & 67    & (78)\\

   \noalign{\smallskip}
   \hline
\end{tabular}
\end{center}
\end{table}

\subsection{The behaviour of the de Jager rate with  effective temperature}

 Figure 6 shows in particular the temperature dependance 
for the van Loon and de Jager prescriptions. It is surprising that for log~$L \ga 4.5$,
the \Mdot\ of de Jager gets smaller for decreasing $T_{\rm eff}$. We cannot explain this 
behaviour in view of the data
concerning the 15 RSGs that dJ88 used when deriving their formula 
(these data are listed in Appendix~A).
In particular,  VY CMa has the largest mass-loss rate of their small sample
($2.4 \times 10^{-4}$\ \Msolyr) and its temperature adopted by dJ88 is low, 
$T_{\rm eff} = 2840$ K. In addition, the weight assigned to this star by dJ88 is
considerably larger (60) than the weight of any other RSG in the sample (weights $\sim$ 1-4).
Yet, the de Jager rate decreases for cooler temperatures.
A possible explanation of this peculiar $T_{\rm eff}$  behaviour  is 
the fact that the fit of the data for the dJ88 rate was made globally on all stars
over the H-R diagram  and not just  ``locally'' on the RSGs.

\subsection{ Is the scatter of \Mdot\  at a given luminosity due to effective temperature ?}

 An interesting question is whether the deviation of the measured \Mdot\   from the dJ88 rate
is correlated with effective temperature. Thanks to the work of LM05, $T_{\rm eff}$ is known 
for the 27 first RSGs of Table 1  (from V589 Cas to TZ Cas). $T_{\rm eff}$ is also known for the 
8 stars of Table 6. We adopt the infrared-based \Mdot\  from Table 1, but prefer when possible
 \Mdot$_{\rm gas}$  from Table 6. Figure~7 (panel a) plots the ratio (in log) of measured
mass-loss rate  to the de Jager rate. It shows a slight tendency of \Mdot\ decreasing
with increasing $T_{\rm eff}$.  The standard deviation is 0.410 when all 31 stars are considered
and it can be seen that 6 stars (of 31) have ratios larger than 4 or smaller than 1/4.
Changing the de Jager rate by adding a $T_{\rm eff}^{-\alpha}$ factor would  slightly 
reduce  this scatter. Values of $\alpha$ between 6 and 10 are suitable.
  We show in panel~b the resulting situation for 
$(T_{\rm eff}/3600)^{-10}$. The standard deviation is 0.367 for all objects included. It is
0.296 if $\mu$ Cep and V396 Cen are omitted, and all other objects are within 
a factor of 4 of agreement. V396 Cen has a poor quality 60~$\mu$m flux, 
coded F in the IRAS point source catalog, and presents an anormally low  
$f_{60}/f_{25}$ flux ratio. 
Therefore its position in Fig.7 is plausibly explained by an underestimated 
60~$\mu$m flux.

\begin{figure}
\includegraphics*[width=8cm,angle=-90]{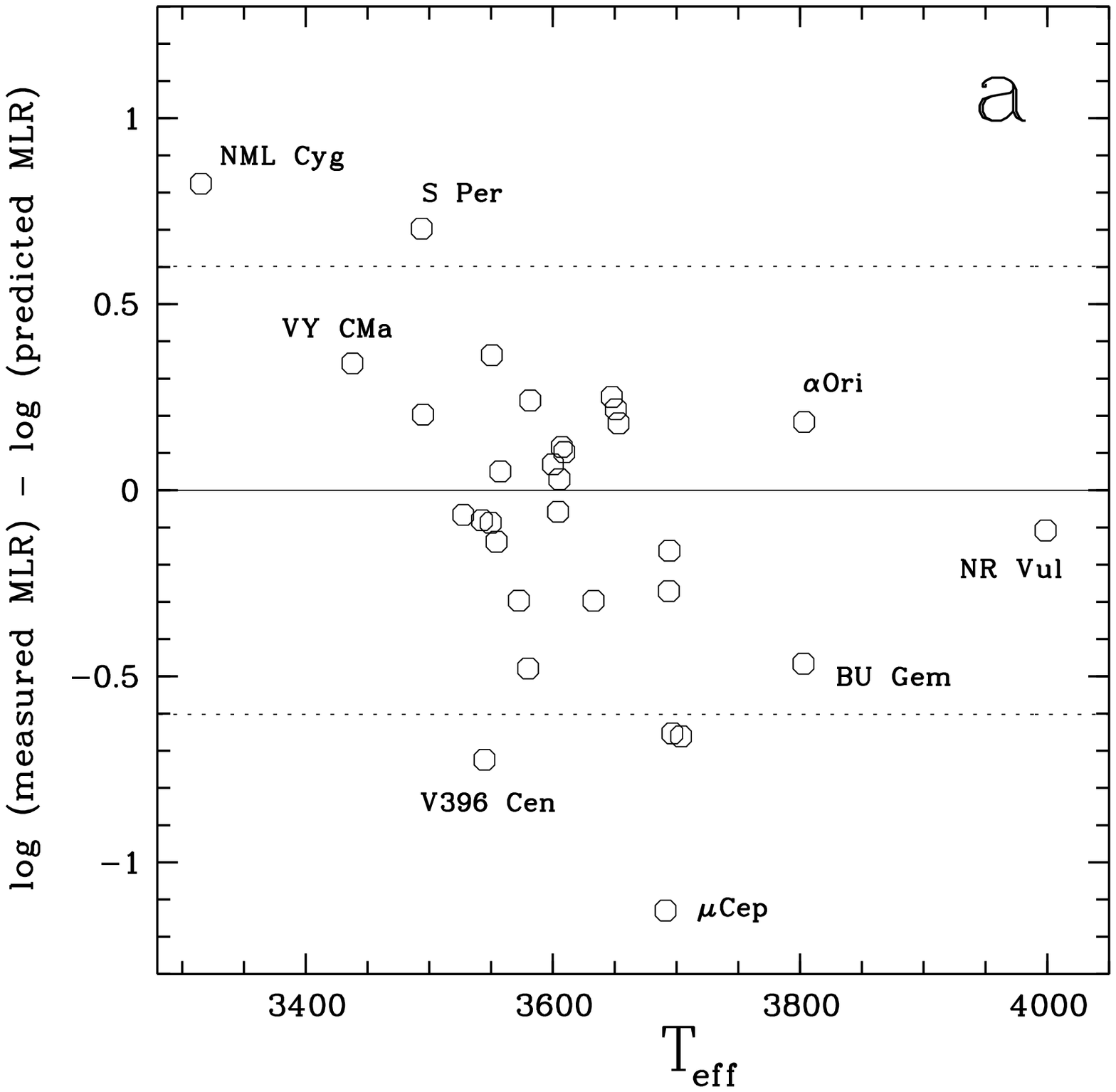}
\includegraphics*[width=8cm,angle=-90]{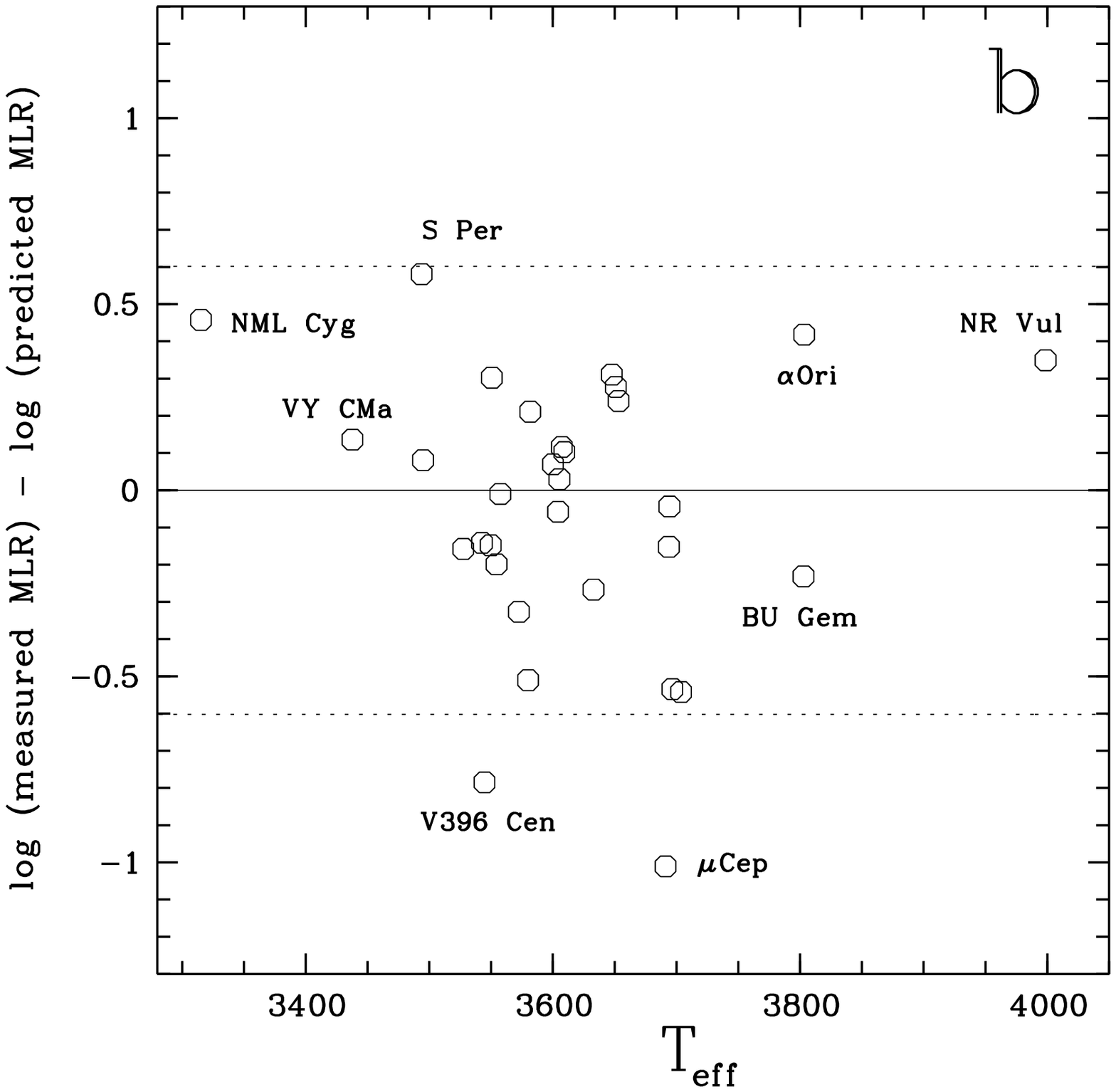}
\caption[]{ {\it Panel a}:  Discrepancies between measured mass-loss rates (MLR)  and
mass-loss rates predicted by the de Jager  prescription, plotted versus
$T_{\rm eff}$, for Galactic red supergiants. The horizontal dotted 
lines indicate discrepancies by factors 4 and 1/4.
{\it Panel b}: same as panel (a) when the de Jager rate is multiplied 
by ($T_{\rm eff}/3600)^{-10}$.}
\end{figure}

\subsection{ The RSGs in the Magellanic Clouds and the dependence on metallicity} 

We have seen that the mass-loss rates in the LMC from Groenewegen et al.\,\,do not
follow fully the de Jager rate. For  $L$ between 63000 and 160\ 000, there is 
reasonable agreement, but at larger $L$ , the de Jager rate is off (too large) 
by a factor of  $\sim$ 8. Recently, Bonanos et al.\,\,(2010) have also considered
 the mass-loss rates of RSGs in the LMC and SMC. Their sample of RSGs ($\sim$ 100 
sources in the LMC, 34 in the SMC) is taken from a list of luminous objects 
having optical spectral classification and photometry from 0.3 to 24 $\mu$m obtained 
during the Spitzer SAGE survey of the Magellanic Clouds (Meixner et al.\,\,2006).
Bonanos et al.\,\,(2010) estimate mass-loss with a $\Mdot$\ versus the $K - [24]$ 
relation derived from model dusty envelopes. 

From the data plotted in their Fig.~17, which show important scatter at a given luminosity, 
 we derive median mass-loss rates in several bins in $L$. 
The 5 bins for the LMC have respectively 15, 24, 28, 25 and 5 data points.
The 4 bins for the SMC have 6, 10, 11 and 7 data points.  The resulting $\Mdot$\ versus 
$L$ relations are shown in Fig.~8 (histograms).  Both the SMC and LMC data suggest 
that $\Mdot$\ increases with luminosity if log\,$L$ is less than $\sim$~5.3, as 
already concluded by Bonanos et al.~(2010). It can be noted that the LMC 
bin with the highest luminosity suggests that $\Mdot$\ could saturate, but the 
low number of points makes it difficult to be sure of that (see these 5 points at
 the highest luminosity in Fig.~17 of Bonanos et al.\,\,2010). 
However, we note that this is quite consistent with our Fig.~5  and its high-$L$, 
low-$\Mdot$\ group of stars, as mentioned above.

 In order to investigate how the RSG mass-loss rates depend on metallicity, we can compare
the median rates obtained above with the dJ88 prescription scaled by various $Z$ values.
 Here we adopt $Z$(LMC) $= Z_{\sun}/2$, and $Z$(SMC) $= Z_{\sun}/5$.  
Figure~8  shows the de Jager lines for warm effective temperatures
which are suitable for the Clouds. It shows that, for the SMC, the observed rates 
are between the de Jager rate scaled by $Z^{0.5}$ and $Z$, and a $Z^{0.7}$ scaling 
would fit the data quite well. As for the LMC, the situation is less clear because 
the metallicity decrease is only a factor 2 with respect to solar. If we omit the most 
luminous bin, a $Z^{0.7}$ scaling is not unreasonable, but a $Z^{0.5}$ is also possible.

 \begin{figure}
\centering
\includegraphics[width=9cm,angle=-90]{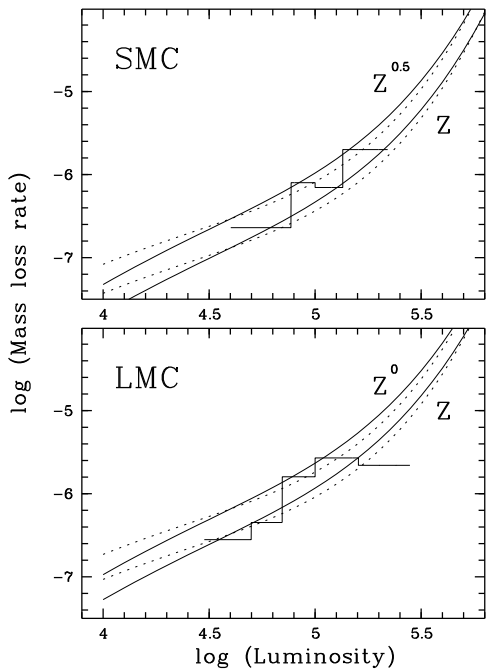}

\caption[]{Mass-loss rates versus luminosity for Magellanic Clouds. 
The (median) data of Bonanos et al.\,\,(2010) are represented by histograms (see text).
The solid and dotted lines show the dJ88 rates for 3500 and 4000\,K
 scaled for various metallicity, as indicated in the panels.}
\end{figure}

 \section{Conclusions}

  The de Jager  prescription  used in evolution codes for RSGs 
  was established in 1988  when the mass-loss rates of stars 
{\it located over all the H-R diagram} were fitted as a function of $L$ and
 $T_{\rm eff}$. The fit was not  specifically made
 for this type of stars, and  the goal of this work was to compare the de Jager rate
 to estimates of $\Mdot$ based on data acquired after 1988. Our main conclusions 
 are the following: 

\smallskip
\smallskip

\noindent 1.~   When the Jura's formula is adopted to get an observational estimate of $\Mdot$\,
   based on the IRAS infrared excess at 60\,$\mu$m and supposing a gas-to-dust ratio of 200 
 (corresponding to complete condensation of refractory material), it is found that the Galactic
   RSGs are satisfactorily distributed around the de Jager line.
    The deviations from de Jager rate
   is in general less than a factor of 4.

\smallskip
\smallskip

\noindent 2.~  Only eight Galactic RSGs have their circumstellar gas detected, i.e. through 
   detection of  \HI\ or CO emission, or other methods permitting to estimate $\Mdot$ from
 observation of the gas.   For these stars,     $\Mdot$  agrees reasonably well
 with the de Jager rate. The most discrepant cases are found at high luminosity,
where the de Jager prescription underestimates $\Mdot$ of NML Cyg by a factor of 7, but
overestimates $\Mdot$ for $\mu$ Cep by a factor of 10 (see Table 3).

\smallskip
\smallskip
 
\noindent 3.~ The Spitzer data from Bonanos et al.\,\,(2010) show that the mass-loss rates 
of RSGs in the Small Magellanic Cloud agree on average with the de Jager rate 
scaled for metallicity by  $(Z/Z_{\sun})^{0.7}$.


\smallskip
\smallskip

\noindent 4.~ A similar behaviour is  less clear with RSGs of the Large Magellanic Cloud. For
luminosities smaller than 1.6 $\times$ 10$^5$ \Lsol, the median mass-loss rates roughly follow
the de Jager prescription and its increasing trend with luminosity. These measured rates
 are also roughly compatible with the $Z^{0.7}$ scaling found for the SMC. 
However, for higher luminosities, one finds that the
mass-loss rates level off at a rate of $\sim$ 3 $\times$ 10$^{-6}$\,\Msolyr\ and are
 smaller by a factor of $\sim$ 8 than the (unscaled) 
dJ88 rate (with the exception of WOH-G64).

\smallskip
\smallskip

\noindent 5.~ Finally, although the dJ88 prescription presents
some odd aspects that can be criticized (e.g., its  behaviour with $T_{\rm eff}$),
it passes rather satisfactorily several tests (listed 1,2,3 and part of 4 above). Therefore,
we think it more appropriate to keep it unchanged in the stellar evolution models
but to apply a $Z^{0.7}$ factor for metallicity dependence.
The LMC mass-loss saturation issue needs to be clarified. We hope to further contribute
to that subject in the future.

\begin{acknowledgements}
We thank our referee, J.Th. van Loon, for many remarks that considerably improved the paper.
N.M. warmly thanks Olivier Richard and Tim Kendall for their help concerning computers 
and english, respectively. N.M. also thanks Raphael Hirschi for discussions about the 
Geneva code. We acknowledge the use of the database produced by the Centre de 
Donn\'ees de Strasbourg (CDS).  This research was partly supported by the French 
National Research Agency (ANR) through program number ANR-06-BLAN-0105.
\end{acknowledgements}


\appendix

\section{List of red supergiants considered by de Jager et al.\,\,(1988)}

The list of RSGs considered by dJ88 are in Table A1. The data from dJ88 are: 
the stellar effective temperature $T_1$ (K), the luminosity $L_1$ (\Lsol)
 and the mass-loss rate $\Mdot_1$ in \Msolyr. The columns $T_2$ , $L_2$  and
$\Mdot_2$  give the same quantities as derived in this work. Many mass-loss rates
used by dJ88 ($\Mdot_1$)  came from the exploitation and analysis of infrared 
flux measurements between 3.5 and 11.5 $\mu$m\ (Gehrz \& Woolf 1971). The mass-loss rates
of Hagen (1982) are derived from flux measurements at 20 and 25 $\mu$m.
The mass-loss rates of Sanner (1976) are based on high resolution spectral profiles of
strong resonance lines. For objects with $\Mdot_1$ based on several individual 
determinations (from 4 to 9 determinations for $\mu$ Cep and  $\alpha$ Ori, 
respectively), see the details in de Jager et al.\,\,(1988).

\begin{table*}[!ht]
\caption[]{List and data for the RSGs considered by dJ88.}
\begin{center}
\begin{tabular}{lrrccccl}
\hline
\hline

Object &  $T_1$  & $T_2$ & log\,$L_1$  & log\,$L_2$ & log($\Mdot_{1}$) & log($\Mdot_{2}$) & Note\\

\hline

\noalign{\smallskip}

TV Gem        & 3560 & 3700 & 4.75  & 4.92  & $-$5.92 & $-$6.46 & GW\\      
BU Gem        & 3440 & 3800 & 5.15  & 4.93  & $-$5.68 & $-$6.23 & GW \\      
$\alpha$ Ori  & 3500 & 3810 & 4.87  & 4.75  & $-$6.02 & $-$5.82 & (1)\\     
$\alpha$ Sco  & 3540 & 3550 & 4.74  & 4.85  & $-$6.08 & $-$6.00 & (1)\\     
RW Cyg        & 3365 & 3600 & 5.20  & 5.16  & $-$5.30 & $-$5.50 & GW\\       
VV Cep        & 3175 & ...  & 5.13  & ...   & $-$6.89 & ...     & GW\\       
CE Tau        & 3400 &  ... & 4.55  & ...   & $-$6.62 & ...     & (4)\\  
$\mu$ Cep     & 3400 & 3700 & 4.55  & 5.61  & $-$5.78 & $-$5.30 & (1)\\     
BC Cyg        & 3155 & 3575 & 5.54  & ...   & $-$5.15 & ...     & GW (2)\\       
BI Cyg        & 3155 & 3575 & 5.38  & 5.09  & $-$5.15 & $-$5.33 & GW\\       
VY CMa        & 2840 & 3430 & 5.76  & 5.47  & $-$3.62 & $-$4.40 & (1)\\     
EV Car        & 2930 & ...  & 5.74  & 5.83  & $-$6.00 & $-$4.90 & (3)\\
S Per         & 2810 & ...  & 5.66  & 4.93  & $-$4.57 & $-$5.16 & GW\\     
UY Sct        & 3000 & ...  & 5.44  & ...   & $-$5.22 & ...     & GW\\ 
AH Sco        & 2800 & ...  & 5.62  & ...   & $-$6.00 & ...     & (3)\\    
NML Cyg       & 2485 & 3309 & 5.80  & 5.50  & $-$4.18 & $-$3.85 & (1)\\     

\noalign{\smallskip}
\hline
\end{tabular}
\end{center}
Notes: GW indicates that $\Mdot_1$ was taken by dJ88 from Gehrz and Woolf~(1971).
 (1) indicates that $\Mdot_1$  is the average of several determinations  obtained with different methods. 
(2) BC Cyg has no IRAS fluxes at  25 and 60~$\mu$m and was not included in this study.
(3)  $\Mdot_1$ from Hagen~(1982). (4) $\Mdot_1$  from Sanner~(1976)

\end{table*}

\section{Photometric data of the studied RSGs}

Table B1 presents the magnitudes in the  UBVIJHK filters and 
the IRAS fluxes at 12, 25, and 60 $\mu$m that were
used to calculate the flux averaged wavelength $\lambda_{\rm m}$, and the luminosities $L$.
In case that a magnitude in the above filters was not available, we used the following relations
which were established from the sample data:\\

$ U \approx 0.62 + 0.78$\,\,\,$(B-V)$\\

$ I \approx V - 0.80$\,\,\,$(V-J)$\\

$ L \approx K-0.27 -0.119$\,\,\,$(K-[12])$\\

$f_{\rm 60} \approx 0.20$\,\,\,$f_{\rm 25}$\\

$f_{100} \approx  0.33$\,\,\,$ f_{60}$\\


\begin{table*}[!ht]
\caption[]{Adopted photometry 
 for calculating $L$ and $\lambda_{\rm m}$. The UBVIJHKL are in magnitudes, and
the fluxes $f$ are in Janskys}
\begin{center}
\begin{tabular}{lrrrrrrrrrrr}
\hline
\hline
\noalign{\smallskip}
Object & U & B & V & I & J & H & K & L & $f_{\rm 12}$ & $f_{\rm 25}$ & $f_{\rm 60}$\\
\noalign{\smallskip}
\hline
\noalign{\smallskip}

BD+60335 = V589 Cas  & 13.94 & 11.49 &  9.15 &  5.03 &  4.06 &  3.01 &  2.49 &  2.16 &   12.74 &    8.67 &    3.61\\
BD+56512 = BU Per   & 14.35 & 11.70 &  9.23 &  4.80 &  3.68 &  2.68 &  2.19 &  1.85 &   44.97 &   31.29 &    5.23\\
HD 14469 = SU Per    & 12.04 &  9.80 &  7.63 &  3.82 &  2.82 &  1.93 &  1.46 &  1.13 &   48.71 &   30.66 &    6.87\\
HD 14488 = RS Per    & 12.90 & 10.62 &  8.35 &  4.33 &  3.05 &  2.11 &  1.56 &  1.22 &   74.44 &   47.82 &    9.93\\
HD 14528 = S Per     & 14.51 & 11.88 &  9.23 &  4.48 &  2.95 &  1.84 &  1.12 &  0.55 &  339.40 &  233.20 &   40.59\\
HD 14826 = V441 Per  & 13.02 & 10.56 &  8.24 &  4.30 &  3.47 &  2.47 &  2.04 &  1.60 &   18.53 &   13.25 &    3.54\\
HD 936979 = YZ Per    & 13.14 & 10.55 &  8.20 &  4.34 &  3.26 &  2.30 &  1.91 &  1.64 &   38.85 &   26.12 &    5.28\\
W Per      & 14.80 & 12.24 &  9.62 &  4.75 &  3.09 &  2.00 &  1.57 &  1.51 &   90.58 &   78.86 &   14.87\\
BD+57647   & 14.93 & 12.26 &  9.52 &  4.76 &  3.83 &  2.71 &  2.11 &  1.56 &   39.04 &   26.45 &    6.47\\
HD 37536 = NO Aur     & 10.52 &  8.30 &  6.21 &  3.01 &  2.12 &  1.13 &  0.97 &  0.51 &   43.49 &   22.89 &    5.12\\
$\alpha$ Ori &  4.36 &  2.29 &  0.40 & -2.47 & -2.93 & -3.73 & -4.01 & -4.43 & 4680.00 & 1740.00 &  299.00\\
HD 42475 = TV Gem   & 10.58 &  8.81 &  6.56 &  3.17 &  2.20 &  1.20 &  0.94 &  0.54 &   96.08 &   41.16 &    6.06\\
HD 42543 = BU Gem  & 11.11 &  8.63 &  6.39 &  3.07 &  2.23 &  1.28 &  0.98 &  0.62 &   78.20 &   47.59 &   10.50\\
CD-314916 = V384 Pup & 13.37 & 11.07 &  8.91 &  5.43 &  4.61 &  3.49 &  3.14 &  2.83 &   10.85 &    7.34 &    2.76\\
HD 90382 = CK Car   & 12.00 &  9.66 &  7.45 &  3.55 &  2.63 &  1.71 &  1.36 &  1.03 &  113.40 &   70.10 &   13.98\\
HD 97671 = V602 Car  & 13.50 & 10.91 &  8.39 &  3.61 &  2.49 &  1.53 &  0.99 &  0.09 &  174.90 &   85.13 &   12.40\\
V396 Cen & 12.30 & 10.00 &  7.85 &  3.58 &  2.58 &  1.56 &  1.20 &  0.79 &   53.30 &   39.81 &    4.98\\
KW Sgr   & 14.81 & 12.50 &  9.66 &  4.23 &  3.12 &  1.99 &  1.54 &  0.91 &  250.10 &  148.30 &   18.39\\
HD 3309034 = NR Vul    & 15.41 & 12.41 &  9.36 &  4.49 &  3.25 &  2.14 &  1.69 &  1.00 &  106.30 &   58.80 &   12.28\\
BD+364025 = BI Cyg    & 14.83 & 12.34 &  9.33 &  3.64 &  2.35 &  1.15 &  0.58 &  0.02 &  334.60 &  244.90 &   51.23\\
KY Cyg    & 17.44 & 14.63 & 11.14 &  4.01 &  2.35 &  0.85 &  0.17 & -0.46 &  510.60 &  329.60 &   50.74\\
BD+394208 = RW Cyg    & 13.43 & 11.01 &  8.13 &  3.38 &  2.11 &  0.94 &  0.45 & -0.08 &  298.40 &  189.80 &   60.69\\
$\mu$ Cep  &  8.88 &  6.43 &  4.17 &  0.31 & -0.52 & -1.30 & -1.65 & -2.09 & 1296.00 &  607.70 &  127.00\\
Case 75 = V 354 Cep  & 16.95 & 13.85 & 10.67 &  5.20 &  3.92 &  2.57 &  1.87 &  1.39 &   77.67 &   46.56 &    8.01\\
Case 78 = V355 Cep   & 15.47 & 13.06 & 10.76 &  5.78 &  4.61 &  3.29 &  2.68 &  2.28 &   19.04 &   14.47 &    3.27\\
BD +602613 = PZ Cas    & 12.84 & 11.48 &  8.90 &  3.90 &  2.42 &  1.53 &  1.02 &  0.39 &  373.00 &  398.20 &   96.48\\
BD +602634 = TZ Cas    & 14.17 & 11.68 &  9.17 &  4.62 &  3.34 &  2.31 &  1.88 &  1.37 &   75.07 &   51.70 &    9.47\\
          &          &       &       &       &       &       &       &       &         &         &\\
EV Car     & 12.43 & 10.09 &  7.89 &  3.21 &  2.11 &  1.21 &  0.79 &  0.32 &  265.40 &  164.00 &   25.87\\
HS Cas     & 14.52 & 12.33 &  9.82 &  5.13 &  3.98 &  2.88 &  2.46 &  2.05 &   23.47 &   16.41 &    3.51\\
XX  Per    & 11.72 & 10.36 &  8.26 &  4.25 &  3.06 &  2.21 &  1.83 &  1.19 &   70.76 &   28.42 &    4.23\\
KK Per     & 12.36 &  9.97 &  7.73 &  4.30 &  3.00 &  2.14 &  1.68 &  1.78 &   19.37 &   10.87 &    2.23\\
AD Per     & 12.66 & 10.13 &  7.85 &  4.30 &  3.38 &  2.48 &  1.94 &  1.65 &   21.28 &   13.90 &    2.85\\
PR Per     & 12.80 & 10.14 &  7.84 &  4.60 &  3.56 &  2.68 &  2.25 &  2.14 &   12.79 &    9.39 &    2.37\\
GP Cas     & 14.67 & 12.08 &  9.37 &  4.76 &  3.53 &  2.38 &  1.95 &  1.51 &   25.73 &   18.47 &    4.45\\
VY CMa     & 12.01 & 10.19 &  7.95 &  3.10 &  1.96 &  0.43 & -0.73 & -2.68 & 9919.00 & 6651.00 & 1453.00\\
$\alpha$ Sco  &  4.08 &  2.78 &  0.96 & -1.92 & -2.55 & -3.47 & -3.70 & -4.04 & 3198.00 &  689.90 &  115.50\\
VX Sgr     & 11.72 &  9.41 &  6.52 &  2.11 &  1.23 &  0.13 & -0.50 & -1.61 & 2738.00 & 1385.00 &  262.70\\
Case 49    & 17.33 & 14.26 & 11.12 &  5.41 &  4.07 &  2.74 &  2.15 &  2.01 &   17.86 &   11.01 &    6.58\\
U Lac      & 12.50 & 11.04 &  8.70 &  4.30 &  2.90 &  2.13 &  1.57 &  0.77 &  124.00 &   61.49 &    9.04\\
 
\noalign{\smallskip}
\hline
\end{tabular}
\end{center}
\smallskip
\end{table*}


\section{Wind expansion velocities}
Figure C.1 shows the wind expansion velocities of several RSGs plotted $versus$ luminosity.
The Galactic RSG data are from Table 2 of this paper. The LMC RSG data are from Marshall 
et al.\,\,(2004) (their Table 1 for luminosities and their Table 2 for expansion velocities).
The error bars represent typical uncertainties on luminosity, $\pm$ 30\% for the Galaxy and
$\pm$20 \% for the LMC. The straight lines have the slope predicted by Habing et al.\,\,(1994)
for dust-driven winds and were used  in the Jura's formula for red supergiants 
lacking a wind velocity measurement.

\begin{figure}
\centering
\includegraphics[height=5cm,angle=-90]{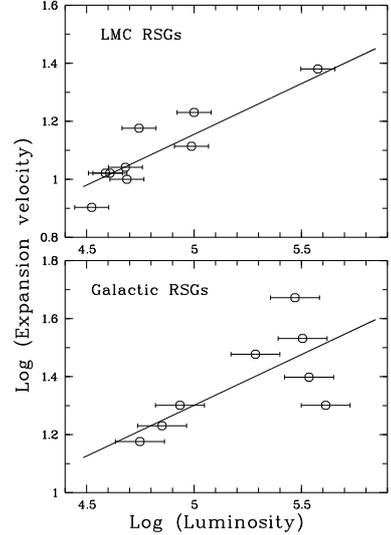}
\caption[]{Wind expansion velocity, in \kms,   as a function of luminosity
(in \Lsol), for Galactic RSGs (lower panel), and LMC RSGs
(upper panel). The straight line for Galactic RSGs corresponds to 
$ V = 20\, (L/10^{5})^{0.35}$ in \kms. The straight line for LMC RSGs corresponds to
  $V = 14\, (L/10^{5})^{0.35}$. }
\end{figure}


\begin{thebibliography}{}
 

\bibitem[2010]{bennett10}
Bennett, P.D. 2010, Chromospheres and Winds of Red Supergiants: An Empirical 
Look at Outer Atmospheric Structure, in Hot and Cool: Bridging Gaps in Massive Star Evolution,
ed. C. Leitherer, P.D. Bennett, P.W. Morris, \& J. van Loon (ASP Conf. Series),
in press; also at astro-ph/1004.1853

\bibitem[1981]{bernat81}
Bernat, A. 1981, ApJ, 246, 184

\bibitem[2001]{blocker01}
Bl\"{o}cker, T., Balega, Y., Hofmann, K.-H., \& Weigelt, G. 2001, A\&A, 369, 142

\bibitem[2009]{bonanos09}
Bonanos, A.Z., Massa, D.L., Sewilo, M., et al.\ 2009, ApJ, 138, 1003

\bibitem[2010]{bonanos10}
Bonanos, A.Z., Lennon, D.J., K\"ohlinger, F., et al. 2010, AJ, 140, 416

\bibitem[1987]{bowers87}
 Bowers, P.F., \& Knapp, G.R. 1987, ApJ, 315, 305 

\bibitem[1989]{cardelli89}
Cardelli, J.A., Clayton, G.C., Mathis, J.S. 1989, ApJ, 345,245

\bibitem[2007]{chen07}
 Chen, X., Shen, Z.-Q., \& Xu, Y. 2007, Chin. J. Astron. Astrophys., 7, 531

\bibitem[2006]{chevalier06}
Chevalier, R.A., Fransson, C., \& Nymark, T.K. 2006, ApJ, 641, 1029
 
\bibitem[2008]{choi08}
Choi, Y.K., Hirota, T., Honma, M., et al.\ 2008, PASJ, 60, 1007

\bibitem[2006]{decin06}
Decin, L., Hont, S., de Koter, A., et al.\ 2006, A\&A, 456, 549
 
 \bibitem[1988]{deJager88}
 de Jager, C., Nieuwenhuijzen, H., \& van der Hucht, K.A. 1988,
 A\&A Suppl. Ser., 72, 259 (dJ88)

\bibitem[2008]{dewit08}
de Wit, W.J., Oudmaijer, R.D., Fujiyoshi, T., et al.\ 2008, ApJ, 685, L75

\bibitem[2008]{eggenberger08}
Eggenberger, P., Meynet, G., Maeder, A., et al.\ 2008, Astrophys.
 Space Science, 316, 43
 
\bibitem[1992]{feast92}
Feast, M.W. 1992, Pulsation and Instability amongst the most Luminous Stars, in
Instabilities in Evolved Super and Hypergiants, ed. C. de Jager,  \& H. Nieuwenhuijzen
(North Holland, Amsterdam), 18

\bibitem[1971]{gehrz91}
Gehrz, R.D., \& Woolf, N.J. 1971, ApJ, 165, 285


\bibitem[1993]{gezari93}
Gezari, D.Y., Schmitz, M., Pitts, P.S., \&  Mead, J.M. 1993,
Catalog of Infrared Observations, Third Edition, NASA Reference Publication 1294


\bibitem[1986]{glassgold86}
 Glassgold, A.E., \& Huggins, P.J. 1986, ApJ, 306, 60

\bibitem[2009]{groenwegen09}
Groenewegen, M.A.T., Sloan, G.C., Soszy\'{n}ski, I., \& Petersen, E.A. 2009, A\&A, 506, 1277

\bibitem[1996]{guilainmauron96}
Guilain, Ch., \& Mauron, N. 1996, A\&A, 314, 585

\bibitem[1997]{gustafsson97}
Gustafsson, B., Eriksson, K., Kiselman, D., Olander, N., \& Olofsson, H. 1997,
A\&A, 318, 535

\bibitem[1994]{habing94}
Habing, H.J.,  Tignon, J., \&  Tielens, A.G.G.M. 1994, A\&A, 286, 523


\bibitem[2008]{harper08}
Harper, G., Brown, A., \& Guinan, E.F. 2008, ApJ, 135, 1430

 
 \bibitem[1994]{huggins94}
 Huggins, P.J., Bachiller, R., Cox, P., \& Forveille, T. 1994, ApJ, 424, L127

  \bibitem[1978]{humphreys78}
 Humphreys, R.M. 1978, ApJSS, 38, 309
 
 \bibitem[2000]{josselin00}
 Josselin, E., Blommaert, J.A.D.L., Groenewegen, M.A.T., Omont, A., 
 \& Li, F.L. 2000, A\&A, 357, 225 (JB00)

\bibitem[2009]{josslancon09}
Josselin, E.,  \& Lan\c{c}on, A. 2009, astro-ph/0911.0840
 
 \bibitem[2007]{josselinplez07}
 Josselin, E., \& Plez, B. 2007, A\&A, 469, 671
 
 \bibitem[1990]{jurak90}
 Jura, M., \& Kleinmann, S.G. 1990, ApJSS, 73, 769 (JK90)

\bibitem[2003]{kemper03}
Kemper, F., Stark, R., Justtanont, K., et al.\ 2003, A\&A, 407, 609

\bibitem[1993]{keene93}
Keene, J., Young, K., Phillips, T.G., \& B\"uttgenbach, Th.H. 1993,
ApJ, 415, L131

\bibitem[2006]{kiss06}
Kiss, L.L., Szab{\'o}, G.M., \& Bedding, T.R. 2006, MNRAS, 372, 1721

\bibitem[1982]{knapp82}
Knapp, G.R., Phillips, T.G., Leighton, R.B., et al.\ 1982, ApJ, 252, 616

\bibitem[1989]{knapp89}
Knapp, G.R., Sutin, B.M., Phillips, T.G., et al.\ 1989, ApJ, 336, 822

\bibitem[1978]{kudritzki78}
Kudritzki, R.P., \& Reimers, D. 1978, A\&A, 70, 227
 
\bibitem[1984]{lambert1984}
Lambert, D.L.,  Brown, J.A., Hinkle, K.H., \& Johnson, H.R. 1984, ApJ, 284, 223

\bibitem[1989]{leborgne89}
Le Borgne, J.-F., \& Mauron, N. 1989, A\&A, 210, 198

\bibitem[1970]{lee70}
 Lee, T.A. 1970, ApJ, 162, 217
 
 \bibitem[2005]{levesque05}
 Levesque, E.M., Massey, P., Olsen, K.A.G., et al.\ 2005, ApJ, 628, 973 (LM05)

\bibitem[2009]{levesque09}
Levesque, E.M., 2009,  The Physical Properties of Red Supergiants, astro-ph/0911.4720,
to appear in New Astronomy Reviews

\bibitem[1989]{maeder89}
Maeder, A., \& Meynet, G. 1989, A\&A, 210,155

\bibitem[2004]{marshall04}
Marshall, J.R., van Loon, J.T., Matsuura, M., et al.\ 2004, MNRAS, 355, 1348


\bibitem[1998]{massey98}
Massey, P. 1998, ApJ, 501,153

\bibitem[2008]{masseyetal2008}
Massey, P., Levesque, E.M., Plez, B., \& Olsen, K.A.G., 2008, astro-ph/0801.1806

\bibitem[2009]{masseyetal2009}
Massey, P., Silva, D.R., Levesque, E.M., et al.\ 2009, ApJ 703, 420

\bibitem[1991]{massey91}
Massey, P., \& Thompson, A.B. 1991, AJ, 101, 1408

\bibitem[1990]{mauron90}
Mauron, N. 1990, A\&A, 227, 141

\bibitem[1996]{meixner96}
Meixner, M., Gordon, K.D., Indebetouw, R., et al. 2006, AJ, 132, 2268
 
 \bibitem[1990]{nieu90}
 Nieuwenhuijzen, H., \& de Jager, C. 1990, A\&A 231, 134

\bibitem[2004]{olofsson04}
Olofsson, H. 2004, Circumstellar Envelopes, in Asymptotic Giant Branch Stars,
ed. Harm J. Habing \& Hans Olofsson (Springer, New-York), 325
 
\bibitem[2002]{plez02}
Plez, B., \& Lambert, D.L. 2002, A\&A, 386, 1009

\bibitem[[2008]{ramstedt08}
Ramstedt, S., Sch\"{o}ier, F.L., Olofsson, H., \& Lundgren, A.A. 2008,
A\&A, 487, 645

\bibitem[1990]{reid90}
Reid, N., Tinney, Ch., \& Mould, J. 1990, ApJ, 348, 98

\bibitem[1975]{reimers75}
Reimers, D. 1975, Mem. Soc. Roy. Sci. Li\`ege, 6$^e$ S\'erie, vol. 8, 369

 \bibitem[2008]{reimers08}
 Reimers, D., Hagen, H.-J., Baade, R., \& Braun, K. 2008, A\&A, 491, 229
 
 \bibitem[1999]{salsnich99}
 Salasnich, B., Bressan, A., \& Chiosi, C. 1999, A\&A, 342, 131

\bibitem[1976]{sanner76}
Sanner, F. 1976, ApJSS, 32, 115
 
\bibitem[2007]{schroder07}
Schr\"{o}der, K.-P., \& Cuntz, M. 2007, A\&A, 465, 593

\bibitem[2006]{schuster06}
Schuster, M.T., Humphreys, R.M., \& Marengo, M. 2006, AJ, 131, 603

\bibitem[2009]{schuster09}
Schuster, M.T., Marengo, M., Hora, J.L., et al.\ 2009, ApJ, 699, 1423

\bibitem[2008]{smartt09}
Smartt, S.J., Eldridge, J.J., Crockett, R.M., \& Maund, J.R. 2009, MNRAS, 395,1409

\bibitem[2009]{smith09}
Smith, N., Hinkle, K., \& Ryde, N. 2009, ApJ 137, 3558

\bibitem[2009]{stancliffe09}
Stancliffe, R.J., \& Eldridge, J.J. 2009, MNRAS, 396,1699
 
\bibitem[1994]{sylvester94}
Sylvester, R.J., Barlow, M.J., \& Skinner, C.J. 1994, MNRAS, 266, 640

\bibitem[1998]{sylvester98}
Sylvester, R.J., Skinner, C.J., \& Barlow, M.J. 1998, MNRAS, 301, 1083

\bibitem[1998]{vanbeveren98}
Vanbeveren, D., De Loore, C., \& Van Rensbergen, W. 1998, A\&A Rev., 9, 63

 \bibitem[2007]{vanbeveren07}
 Vanbeveren, D., Van Bever, J., \& Belkus, H. 2007, ApJ, 662, L107
 
 \bibitem[2005]{vanloon05}
 van Loon, J.Th., Cioni, M.-R. L., Zijlstra, A.A., \& Loup, C. 2005, A\&A, 438, 273

\bibitem[2010]{vanloon10}
van Loon, J.T. 2010, The effects of red supergiant mass loss on supernova 
ejecta and the circumburst medium, in Hot and Cool: Bridging Gaps in 
Massive Star Evolution, ed. C. Leitherer, P.D. Bennett, P.W. Morris, \& J.Th. 
van Loon  (ASP Conf. Series, vol. 425, San Francisco)

\bibitem[2009]{verhoelst09}
Verhoelst T., van der Zypen, N., Hony, S., et al.\ 2009, A\&A 498, 127

\bibitem[2007]{woitke07}
Woitke, P. 2007, What Drives the Mass Loss of Oxygen-Rich AGB Stars, in
Why Galaxies Care about AGB Stars: Their Importance as Actors and Probes,
ed. K. Kerschbaum, C. Charbonnel, \& R.F. Wing (ASP Conf Series, vol. 378, 
San Francisco), 156

\bibitem[1993]{woodhams93}
Woodhams, M. 1993, Mass Loss from Late Type Supergiants, in Massive Stars: 
Their Lives in the Interstellar Medium, ed. J.P. Cassinelli \& E.B. Churchwell,
(ASP Conf. Series, vol. 35, San Francisco), 231
 
 

\end{thebibliography}
\end{document}